\documentclass[zpreprint,zbstpl,british]{zeus_K2.9paper}
\usepackage{graphicx}
\usepackage{booktabs}
\usepackage{multirow}
\usepackage{zeusdet_hepunits}
\usepackage{lineno}%./LaTeX/styles/lineno
\usepackage{array, tabularx, booktabs,longtable}
\def\citeCTD{{\cite{%
nim:a279:290,*npps:b32:181,*nim:a338:254%
}}\xspace}
\def\citeMVD{{\cite{%
nim:a581:656%
}}\xspace}
\def\citeSTT{{\cite{%
nim:a535:191%
}}\xspace}
\def\citeCAL{{\cite{%
nim:a309:77,*nim:a309:101,*nim:a321:356,*nim:a336:23%
}}\xspace}
\def\citeHES{{\cite{%
nim:a277:176%
}}\xspace}
\def\citeSixmT{{\cite{%
thesis:gosau:2007%
}}\xspace}
\def\citeBAC{{\cite{%
nim:a300:480%
}}\xspace}
\def\citePCAL{{\cite{%
desy-92-066,*zfp:c63:391,*acpp:b32:2025%
}}\xspace}
\def\citeSPECTRO{{\cite{%
nim:a565:572%
}}\xspace}

\newcommand{\PYTHIA}{\textsc{Pythia}\xspace}

\newcommand{\Ejet}{\ensuremath{E^{\text{jet}}}\xspace}
\newcommand{\ETjet}{\ensuremath{E_{T}^{\text{jet}}}\xspace}

\newcommand{\ETgam}{\ensuremath{E_{T}^{\gamma}}\xspace}
\newcommand{\etagam}{\ensuremath{\eta^{\gamma}}\xspace}
\newcommand{\etajet}{\ensuremath{\eta^{\text{jet}}}\xspace}
\newcommand{\phijet}{\ensuremath{\phi^{\text{jet}}}\xspace}
\newcommand{\Detaegamma}{\ensuremath{\Delta\eta^{e, \gamma}}\xspace}
\newcommand{\Dphiegamma}{\ensuremath{\Delta\phi^{e, \gamma}}\xspace}

\newcommand{\xgamma}{\ensuremath{x_{\gamma}^{\mathrm{meas}}}\xspace}
\newcommand{\xp}{\ensuremath{x_p^{\mathrm{obs}}}\xspace}
\newcommand{\pzjet}{\ensuremath{p_Z^{\mathrm{jet}}}\xspace}

\newcommand{\yJB}{\ensuremath{y_{\mathrm{JB}}}\xspace}
\newcommand{\zdet}{\footnote{%
The ZEUS coordinate system is a right-handed Cartesian system, with the $Z$ 
axis pointing in the nominal proton beam direction, referred to as the ``forward
direction'', and the $X$ axis pointing  towards the centre of HERA.
The coordinate origin is at the centre of the central tracking detector.
The pseudorapidity is defined as $\eta=-\ln\left(\tan\frac{\theta}{2}\right)$, 
where the polar angle, $\theta$, is measured with respect to the
$Z$ axis. The azimuthal angle, $\phi$, is
measured with respect to the $X$ axis.\xspace}}

\usepackage{array}
\newcolumntype{L}[1]{>{\raggedright\let\newline\\\arraybackslash\hspace{0pt}}m{#1}}
\newcolumntype{C}[1]{>{\centering\let\newline\\\arraybackslash\hspace{0pt}}m{#1}}
\newcolumntype{R}[1]{>{\raggedleft\let\newline\\\arraybackslash\hspace{0pt}}m{#1}}

\usepackage{dcolumn}
\usepackage{color,soul} %to highlight text
\usepackage{color}
\newcolumntype{d}[1]{D{.}{.}{#1} }%start tables 
\begin{document}
%------------------------------------------------------------------------------
%       Title sheet
%------------------------------------------------------------------------------
%
%
%
%\prepnum{ZEUS-prel-16-xxxx}
\prepnum{DESY-17-212}
\prepdate{\today}                     %MMMM YYYY

\title{\Large Further studies of isolated photon production with a jet in deep inelastic scattering at HERA}
                  
\author{ZEUS Collaboration}
\date{}              % The line is needed.  Add parameter \today for date.
\draftversion{}      % It seems the line is needed even if empty

\maketitle

%
% If you use the package units instead of hepunits you have to enclose
% the values in square brackets and change \invpb to \pbi and \GeV to \Gev
% e.g. \unit[47.7]{\pbi}
% If the quantity is not in math mode and the unit contains math mode
% characters such as superscripts it must be contained in $...$
%
\begin{abstract}\noindent
{Isolated photons with high transverse energy have been studied in
deep inelastic $ep$ scattering with the ZEUS detector at HERA, using
an integrated luminosity of $326\, \mathrm{pb}^{-1}$ in the range of
exchanged-photon virtuality $10 - 350 \,\, \mathrm{GeV}^2$. Outgoing
isolated photons with transverse energy $4<\ETgam< 15$ GeV and
pseudorapidity $-0.7 <\etagam< 0.9$ were measured with accompanying
jets having transverse energy and pseudorapidity $2.5 <\ETjet<35$ GeV
and $-1.5<\etajet< 1.8$, respectively.  Differential cross sections
are presented for the following variables: the fraction of the
incoming photon energy and momentum that is transferred to the
outgoing photon and the leading jet; the fraction of the incoming
proton energy transferred to the photon and leading jet; the differences in
azimuthal angle and pseudorapidity between the outgoing photon and the
leading jet and between the outgoing photon and the scattered
electron. Comparisons are made with theoretical predictions: a
leading-logarithm Monte Carlo simulation, a next-to-leading-order QCD
prediction, and a prediction using the $k_T$-factorisation approach. }
\end{abstract}

\thispagestyle{empty}
%------------------------------------------------------------------------------
%       Authors - you may have to play with \clearpage and \cleardoublepage 
%       in order to get the main text to start on the correct page
%------------------------------------------------------------------------------
\clearpage

%===================================================================
%
%  MEMBER NAME  AUTH183 (ZEUS)     M  TEX
%
%  JH.: transformed to a format, which is suited as input for
%       CONVERT, which automatically creates author-indices
%
%  Don't remove lines starting with a percent sign %,
%  CONVERT may need them urgently !
%  
%=====================================================================

                                                   %
\begin{center}
{                      \Large  The ZEUS Collaboration              }
\end{center}

{\small\raggedright

%  members:

H.~Abramowicz$^{24, p}$, 
I.~Abt$^{19}$, 
L.~Adamczyk$^{7}$, 
M.~Adamus$^{30}$, 
R. Aggarwal$^{3, b}$, 
S.~Antonelli$^{1}$, 
V.~Aushev$^{16}$, 
Y.~Aushev$^{16}$, 
O.~Behnke$^{9}$, 
U.~Behrens$^{9}$
% ...
\newpage

%       institutes:

{\setlength{\parskip}{0.4em}
\makebox[3ex]{$^{1}$}
\begin{minipage}[t]{14cm}
{\it INFN Bologna, Bologna, Italy}~$^{A}$

\end{minipage}

\makebox[3ex]{$^{2}$}
\begin{minipage}[t]{14cm}
{\it Physikalisches Institut der Universit\"at Bonn,
Bonn, Germany}~$^{B}$

\end{minipage}

\makebox[3ex]{$^{3}$}
\begin{minipage}[t]{14cm}
{\it Panjab University, Department of Physics, Chandigarh, India}

\end{minipage}

\makebox[3ex]{$^{4}$}
\begin{minipage}[t]{14cm}
{\it Calabria University,
Physics Department and INFN, Cosenza, Italy}~$^{A}$

\end{minipage}

\makebox[3ex]{$^{5}$}
\begin{minipage}[t]{14cm}
{\it National Centre for Particle Physics, Universiti Malaya, 50603 Kuala Lumpur, Malaysia}~$^{C}$

\end{minipage}

\makebox[3ex]{$^{6}$}
\begin{minipage}[t]{14cm}
{\it The Henryk Niewodniczanski Institute of Nuclear Physics, Polish Academy of \\
Sciences, Krakow, Poland}

\end{minipage}

\makebox[3ex]{$^{7}$}
\begin{minipage}[t]{14cm}
{\it AGH University of Science and Technology, Faculty of Physics and Applied Computer
Science, Krakow, Poland}

\end{minipage}

\makebox[3ex]{$^{8}$}
\begin{minipage}[t]{14cm}
{\it Department of Physics, Jagellonian University, Krakow, Poland}

\end{minipage}

\makebox[3ex]{$^{9}$}
\begin{minipage}[t]{14cm}
{\it Deutsches Elektronen-Synchrotron DESY, Hamburg, Germany}

\end{minipage}

\makebox[3ex]{$^{10}$}
\begin{minipage}[t]{14cm}
{\it Deutsches Elektronen-Synchrotron DESY, Zeuthen, Germany}

\end{minipage}

\makebox[3ex]{$^{11}$}
\begin{minipage}[t]{14cm}
{\it School of Physics and Astronomy, University of Glasgow,
Glasgow, United Kingdom}~$^{D}$

\end{minipage}

\makebox[3ex]{$^{12}$}
\begin{minipage}[t]{14cm}
{\it Hamburg University, Institute of Experimental Physics, Hamburg,
Germany}~$^{E}$

\end{minipage}

\makebox[3ex]{$^{13}$}
\begin{minipage}[t]{14cm}
{\it Institute of Particle and Nuclear Studies, KEK,
Tsukuba, Japan}~$^{F}$

\end{minipage}

\makebox[3ex]{$^{14}$}
\begin{minipage}[t]{14cm}
{\it Institute of Physics and Technology of Ministry of Education and
Science of Kazakhstan, Almaty, Kazakhstan}

\end{minipage}

\makebox[3ex]{$^{15}$}
\begin{minipage}[t]{14cm}
{\it Institute for Nuclear Research, National Academy of Sciences, Kyiv, Ukraine}

\end{minipage}

\makebox[3ex]{$^{16}$}
\begin{minipage}[t]{14cm}
{\it Department of Nuclear Physics, National Taras Shevchenko University of Kyiv, Kyiv, Ukraine}

\end{minipage}

\makebox[3ex]{$^{17}$}
\begin{minipage}[t]{14cm}
{\it Meiji Gakuin University, Faculty of General Education,
Yokohama, Japan}~$^{F}$

\end{minipage}

\makebox[3ex]{$^{18}$}
\begin{minipage}[t]{14cm}
{\it Lomonosov Moscow State University, Skobeltsyn Institute of Nuclear Physics,
Moscow, Russia}~$^{G}$

\end{minipage}

\makebox[3ex]{$^{19}$}
\begin{minipage}[t]{14cm}
{\it Max-Planck-Institut f\"ur Physik, M\"unchen, Germany}

\end{minipage}

\makebox[3ex]{$^{20}$}
\begin{minipage}[t]{14cm}
{\it Department of Physics, University of Oxford,
Oxford, United Kingdom}~$^{D}$

\end{minipage}

\makebox[3ex]{$^{21}$}
\begin{minipage}[t]{14cm}
{\it INFN Padova, Padova, Italy}~$^{A}$

\end{minipage}

\makebox[3ex]{$^{22}$}
\begin{minipage}[t]{14cm}
{\it Dipartimento di Fisica e Astronomia dell' Universit\`a and INFN,
Padova, Italy}~$^{A}$

\end{minipage}

\makebox[3ex]{$^{23}$}
\begin{minipage}[t]{14cm}
{\it Polytechnic University, Tokyo, Japan}~$^{F}$

\end{minipage}

\makebox[3ex]{$^{24}$}
\begin{minipage}[t]{14cm}
{\it Raymond and Beverly Sackler Faculty of Exact Sciences, School of Physics, \\
Tel Aviv University, Tel Aviv, Israel}~$^{H}$

\end{minipage}

\makebox[3ex]{$^{25}$}
\begin{minipage}[t]{14cm}
{\it Department of Physics, Tokyo Institute of Technology,
Tokyo, Japan}~$^{F}$

\end{minipage}

\makebox[3ex]{$^{26}$}
\begin{minipage}[t]{14cm}
{\it Universit\`a di Torino and INFN, Torino, Italy}~$^{A}$

\end{minipage}

\makebox[3ex]{$^{27}$}
\begin{minipage}[t]{14cm}
{\it Universit\`a del Piemonte Orientale, Novara, and INFN, Torino,
Italy}~$^{A}$

\end{minipage}

\makebox[3ex]{$^{28}$}
\begin{minipage}[t]{14cm}
{\it Physics and Astronomy Department, University College London,
London, United Kingdom}~$^{D}$

\end{minipage}

\makebox[3ex]{$^{29}$}
\begin{minipage}[t]{14cm}
{\it Faculty of Physics, University of Warsaw, Warsaw, Poland}

\end{minipage}

\makebox[3ex]{$^{30}$}
\begin{minipage}[t]{14cm}
{\it National Centre for Nuclear Research, Warsaw, Poland}

\end{minipage}

\makebox[3ex]{$^{31}$}
\begin{minipage}[t]{14cm}
{\it Department of Particle Physics and Astrophysics, Weizmann
Institute, Rehovot, Israel}

\end{minipage}

\makebox[3ex]{$^{32}$}
\begin{minipage}[t]{14cm}
{\it Department of Physics, York University, Ontario, Canada M3J 1P3}~$^{I}$

\end{minipage}

}

\vspace{3em}

%  references concerning institutes;

{\setlength{\parskip}{0.4em}\raggedright
\makebox[3ex]{$^{ A}$}
\begin{minipage}[t]{14cm}
 supported by the Italian National Institute for Nuclear Physics (INFN) \
\end{minipage}

\makebox[3ex]{$^{ B}$}
\begin{minipage}[t]{14cm}
 supported by the German Federal Ministry for Education and Research (BMBF), under
 contract No.\ 05 H09PDF\
\end{minipage}

\makebox[3ex]{$^{ C}$}
\begin{minipage}[t]{14cm}
 supported by HIR grant UM.C/625/1/HIR/149 and UMRG grants RU006-2013, RP012A-13AFR and RP012B-13AFR from
 Universiti Malaya, and ERGS grant ER004-2012A from the Ministry of Education, Malaysia\
\end{minipage}

\makebox[3ex]{$^{ D}$}
\begin{minipage}[t]{14cm}
 supported by the Science and Technology Facilities Council, UK\
\end{minipage}

\makebox[3ex]{$^{ E}$}
\begin{minipage}[t]{14cm}
 supported by the German Federal Ministry for Education and Research (BMBF), under
 contract No.\ 05h09GUF, and the SFB 676 of the Deutsche Forschungsgemeinschaft (DFG) \
\end{minipage}

\makebox[3ex]{$^{ F}$}
\begin{minipage}[t]{14cm}
 supported by the Japanese Ministry of Education, Culture, Sports, Science and Technology
 (MEXT) and its grants for Scientific Research\
\end{minipage}

\makebox[3ex]{$^{ G}$}
\begin{minipage}[t]{14cm}
 partially supported by RF Presidential grant NSh-7989.2016.2\
\end{minipage}

\makebox[3ex]{$^{ H}$}
\begin{minipage}[t]{14cm}
 supported by the Israel Science Foundation\
\end{minipage}

\makebox[3ex]{$^{ I}$}
\begin{minipage}[t]{14cm}
 supported by the Natural Sciences and Engineering Research Council of Canada (NSERC) \
\end{minipage}

}

\pagebreak[4]
{\setlength{\parskip}{0.4em}

%  references concerning members;

\makebox[3ex]{$^{ a}$}
\begin{minipage}[t]{14cm}
now at INFN Roma, Italy\
\end{minipage}

\makebox[3ex]{$^{ b}$}
\begin{minipage}[t]{14cm}
now at DST-Inspire Faculty, Pune University, India\
\end{minipage}

\makebox[3ex]{$^{ c}$}
\begin{minipage}[t]{14cm}
now at Sant Longowal Institute of Engineering and Technology, Longowal, Punjab, India\
\end{minipage}

\makebox[3ex]{$^{ d}$}
\begin{minipage}[t]{14cm}
also at Agensi Nuklear Malaysia, 43000 Kajang, Bangi, Malaysia\
\end{minipage}

\makebox[3ex]{$^{ e}$}
\begin{minipage}[t]{14cm}
now at Rockefeller University, New York, NY 10065, USA\
\end{minipage}

\makebox[3ex]{$^{ f}$}
\begin{minipage}[t]{14cm}
now at University of Liverpool, United Kingdom\
\end{minipage}

\makebox[3ex]{$^{ g}$}
\begin{minipage}[t]{14cm}
now at Tel Aviv University, Israel\
\end{minipage}

\makebox[3ex]{$^{ h}$}
\begin{minipage}[t]{14cm}
now at Physikalisches Institut, Universit\"{a}t Heidelberg, Germany\
\end{minipage}

\makebox[3ex]{$^{ i}$}
\begin{minipage}[t]{14cm}
supported by the Alexander von Humboldt Foundation\
\end{minipage}

\makebox[3ex]{$^{ j}$}
\begin{minipage}[t]{14cm}
Alexander von Humboldt Professor; also at DESY and University of Oxford\
\end{minipage}

\makebox[3ex]{$^{ k}$}
\begin{minipage}[t]{14cm}
also at DESY\
\end{minipage}

\makebox[3ex]{$^{ l}$}
\begin{minipage}[t]{14cm}
now at European X-ray Free-Electron Laser facility GmbH, Hamburg, Germany\
\end{minipage}

\makebox[3ex]{$^{ m}$}
\begin{minipage}[t]{14cm}
now at Kobe University, Japan\
\end{minipage}

\makebox[3ex]{$^{ n}$}
\begin{minipage}[t]{14cm}
now at RWTH Aachen, Germany\
\end{minipage}

\makebox[3ex]{$^{ o}$}
\begin{minipage}[t]{14cm}
also at University of Bristol, United Kingdom\
\end{minipage}

\makebox[3ex]{$^{ p}$}
\begin{minipage}[t]{14cm}
also at Max Planck Institute for Physics, Munich, Germany, External Scientific Member\
\end{minipage}

\makebox[3ex]{$^{ q}$}
\begin{minipage}[t]{14cm}
now at University of Bath, United Kingdom\
\end{minipage}

\makebox[3ex]{$^{ r}$}
\begin{minipage}[t]{14cm}
also supported by DESY and the Alexander von Humboldt Foundation\
\end{minipage}

\makebox[3ex]{$^{ s}$}
\begin{minipage}[t]{14cm}
also at \L\'{o}d\'{z} University, Poland\
\end{minipage}

\makebox[3ex]{$^{ t}$}
\begin{minipage}[t]{14cm}
member of \L\'{o}d\'{z} University, Poland\
\end{minipage}

\makebox[3ex]{$^{ u}$}
\begin{minipage}[t]{14cm}
now at Polish Air Force Academy in Deblin\
\end{minipage}

}

}

\clearpage
\pagenumbering{arabic}

\section{Introduction}
\label{sec-int}

The isolated high-energy photons that are emitted in high-energy
collisions involving hadrons are predominantly unaffected by parton
hadronisation.  Their production probes the underlying partonic
process and can provide information on the structure of the proton.
Processes of this type have been studied in a number of fixed-target
and hadron-collider experiments
\cite{zfp:c13:207,*zfp:c38:371,*pr:d48:5,*prl:73:2662,*prl:95:022003,*prl:84:2786,*plb:639:151,*atlas1,*atlas2,*cms1}.
The production of isolated photons in photoproduction, where the
incoming photon is quasi-real, was previously studied at HERA by the
ZEUS and H1 collaborations
\cite{pl:b413:201,*pl:b472:175,*pl:b511:19,epj:c49:511,epj:c38:437}.
Deep inelastic neutral current (NC) $ep$ scattering (DIS), in which
the exchanged photon has virtuality $Q^2 > 1\,\, \GeV^{2}$, has also
been measured in a variety of $Q^2$
ranges~\cite{pl:b595:86,pl:b687:16,epj:c54:371}.  The ana\-ly\-sis
presented here extends an earlier ZEUS measurement of isolated photons
and jets in DIS~\cite{pl:b715:88}.

\begin{figure}[p]
\vfill
\begin{center}
\includegraphics[width=0.4\linewidth]{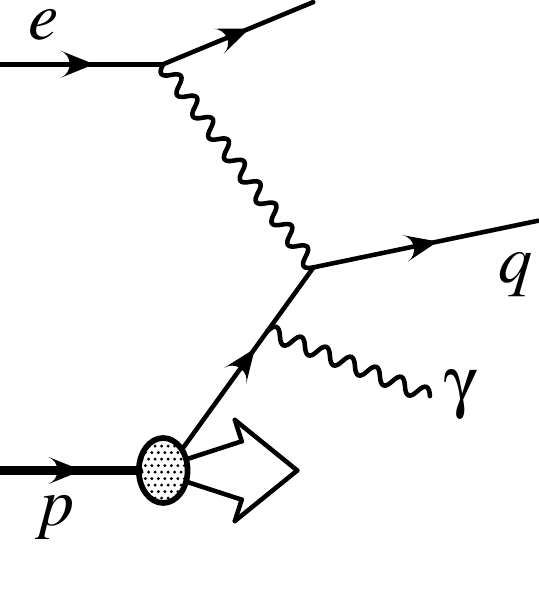}
\includegraphics[width=0.4\linewidth]{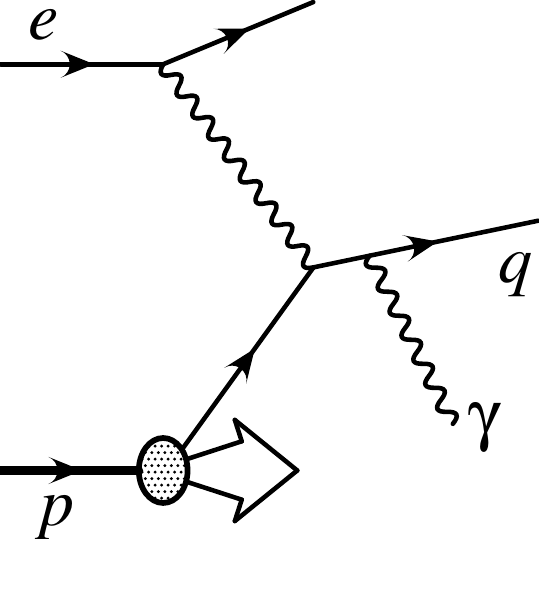}\\
\hspace*{0.1\textwidth}(a)\hspace*{0.48\textwidth}(b)\\[0.04\textwidth]
\vspace{1.5cm}
\includegraphics[width=0.4\linewidth]{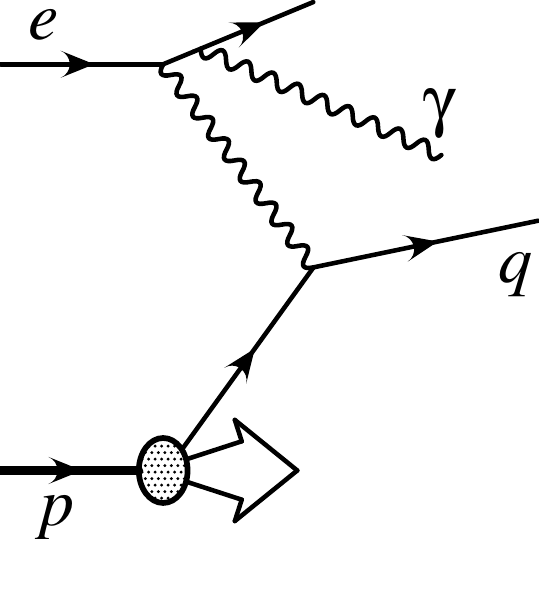}
\includegraphics[width=0.4\linewidth]{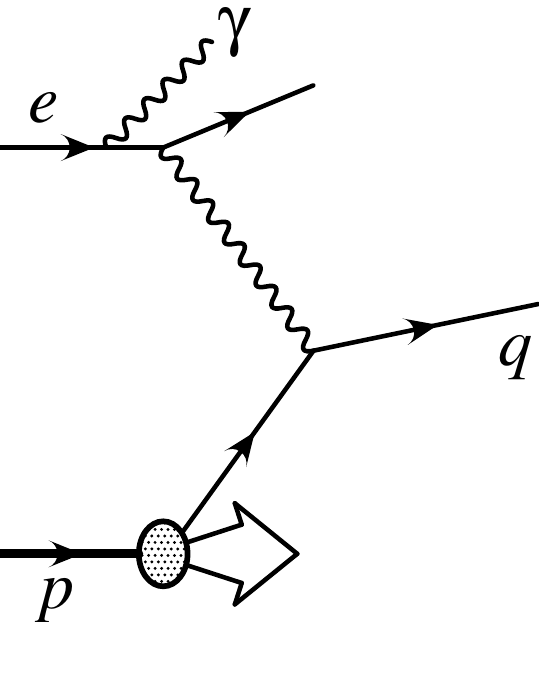}\\
\hspace*{0.1\textwidth}(c)\hspace*{0.48\textwidth}(d)\\[0.04\textwidth]
\vspace{1.5cm}
\end{center}
\caption{\small Lowest-order diagrams for  photon 
production in 
$ep$ scattering.  
(a), (b):  quark radiative diagrams (QQ); (c), (d): lepton radiative diagrams (LL).
\label{fig1}}
\vfill
\end{figure}

Figure~\ref{fig1} shows leading-order diagrams for high-energy photon
production in DIS. Such ``prompt'' photons are emitted either by the
incoming or outgoing quark or by the incoming or outgoing lepton. In
the first case, the photons are classified as ``QQ'' photons, and the
hadronic process has two hard scales: the virtuality $Q^2$ of the
incident exchanged photon and the square of the transverse momentum of
the prompt photon.  In the second case, the photons are denoted as
``LL'' and are emitted from the incoming or outgoing lepton.  The
present analysis requires the observation of a scattered electron, a
high-energy outgoing photon and a hadronic jet. Processes in which the
final state consists solely of a hard outgoing electron and a hard
outgoing photon are thereby excluded.  By requiring the outgoing
photon to be isolated, a further class of processes in which the
photon is produced within a jet is suppressed.

In the previous ZEUS publication on this topic~\cite{pl:b715:88},
kinematic distributions of the outgoing photon and the jet were
studied. Using the same data set, the analysis is extended here by
measuring variables that involve two of the outgoing photon, the jet and the
scattered electron.  Results from a leading-logarithm parton-shower
Monte Carlo~\cite{jhep:0605:026} are compared to the measurements.
Comparison is also made with two theoretical models: one at
next-to-leading order (NLO) in QCD~\cite{epj:c44:395,af}, and one
based on a $k_T$-factorisation approach~\cite{pr:d81:094034}.

\section{Experimental set-up}
\label{sec-exp}
The data sample used for the measurement corresponds to an integrated
luminosity of $326\pm6\,\mathrm{pb}^{-1}$ and was taken with the ZEUS
detector in the years 2004--2007. During this period, HERA ran with an
electron/positron beam energy of 27.5 \GeV\ and a proton beam energy
of 920 \GeV; $138\pm2 \,\mathrm{pb}^{-1}$ of
$e^+p$ data and $188\pm 3 \,\mathrm{pb}^{-1}$ of $e^-p$
data\footnote{Hereafter, ``electron'' refers to both electrons and
positrons unless otherwise stated.} were used in the present analysis.

A detailed description of the ZEUS detector can be found
elsewhere~\cite{zeus:1993:bluebook}. Charged particles were
recorded in the central tracking detector (CTD)~\citeCTD and a silicon
microvertex detector~\cite{nim:a581:656} which operated in a magnetic
field of $1.43$~T provided by a thin superconducting solenoid.  The
high-resolution uranium--scintillator calorimeter (CAL)~\citeCAL
consisted of three parts: the forward (FCAL), the barrel (BCAL) and
the rear (RCAL) calorimeters. The BCAL covered the pseudorapidity
range $-0.74$ to $1.01$ as seen from the nominal interaction
point\zdet. The FCAL and RCAL extended the range to $-3.5$ to $4.0$.  The
smallest subdivision of the CAL is called a cell. The barrel
electromagnetic calorimeter (BEMC) cells had a pointing geometry aimed
at the nominal interaction point, with a cross section approximately
$5\times20\,\mathrm{cm^2}$, with the finer granularity in the
$Z$-direction.  This fine granularity allows the use of shower-shape
distributions to distinguish isolated photons from the products of
neutral meson decays such as $\pi^0 \rightarrow \gamma \gamma$.

The luminosity was measured using the Bethe--Heitler reaction $ep
\rightarrow e\gamma p$ by a luminosity detector which consisted of two
independent systems: a lead--scintillator calorimeter
\cite{desy-92-066,*zfp:c63:391,*acpp:b32:2025} and a magnetic
spectrometer~\cite{nim:a565:572}.

\section{Event selection and reconstruction}
\label{sec-selec}

The ZEUS experiment operated a three-level trigger
system~\cite{zeus:1993:bluebook,uproc:chep:1992:222,nim:a580:1257}.
At the first level, events were selected if they had an energy deposit
in the CAL consistent with an isolated electron. At the second level,
a requirement on the energy and longitudinal momentum of the event was
used to select NC DIS events. At the third level, the full event was
reconstructed and tighter requirements for a DIS electron were made.
Offline selections, similar to those of the earlier ZEUS
analysis~\cite{pl:b715:88}, were then applied.

Outgoing electrons were selected with polar angle \mbox{$\theta_e >
140^{\circ} $} in order to provide a good measurement in the RCAL,
kinematically separated from the selected outgoing photons. Their
impact point ($X$,$Y$) on the surface of the RCAL was required to lie
outside a rectangular region $\pm 14.8$~cm in $X$ and $[-14.6, +12.5]$
cm in $Y$, to give a well understood acceptance.  The outgoing
electrons were identified using a neural network~\cite{sinistra}, and
the energy of the outgoing electron, $E'_e$, corrected for apparatus
effects, was required to be larger than 10 \gev.  The kinematic
variable $Q^2$ was reconstructed as $Q^2=-(k-k')^2,$ where $k$ ($k'$)
is the four-momentum of the incoming (outgoing) electron.  The
kinematic region $10 <Q^2< 350$~\gev$^2$ was selected.
 	
A requirement that the event vertex position, $Z_{\mathrm{vtx}}$,
should be within the range $|Z_{\mathrm{vtx}}|< 40\,\mathrm{cm}$
reduces the background from non-$ep$ collisions. A further requirement
for a well-contained DIS event, $35 < E-p_Z< 65\,\mathrm{\GeV}$, was
imposed where $E-p_Z = \sum \limits_i E_i(1-\cos \theta_i)$; $E_i$ is
the energy of the $i$-th CAL cell, $\theta_i$ is its polar angle and
the sum runs over all cells~\cite{pl:b303:183}.
 	
Photon candidates were identified as energy-flow objects
(EFOs)\footnote{Energy-flow
objects~\cite{epj:c1:81,*epj:c6:43,*briskin:phd:1998} were constructed
from calorimeter-cell clusters and tracks, associated when possible.}
without an associated track, for which at least $90\%$ of the
reconstructed energy was deposited in the BEMC. The calibration of the
energies of the photon and scattered electron was taken from an
earlier ZEUS analysis and used deeply virtual Compton scattering
events~\cite{jhep:08:23}.  The reconstructed transverse energy of the
photon candidate, $E_T^{\gamma}$, was required to lie within the
range\footnote{The upper limit was selected to retain distinguishable
shower shapes between the hadronic background and the photon signal.}
\mbox{$4<E_T^{\gamma}<15\,\mathrm{\GeV}$} and the pseudorapidity,
$\eta^{\gamma}$, had to satisfy $-0.7 < \eta^{\gamma} < 0.9$.

Jets were reconstructed with the $k_T$ clustering
algorithm~\cite{np:b406:187} in the $E$ scheme in the longitudinally
invariant inclusive mode~\cite{pr:d48:3160} with the $R$ parameter set
to 1.0. Since all EFOs of the event were used except for the electron
signal, one of the jets found by this procedure corresponds to or
includes the photon candidate. At least one accompanying jet was
required with transverse energy $\ETjet > 2.5 \,\, \mathrm{GeV}$ and
pseudorapidity, \etajet, in the range $-1.5<\etajet < 1.8$; if more
than one jet was found, that with the highest \ETjet was used.
 	
Photons radiated from final-state electrons were suppressed by
requiring that $\Delta R>0.2$, where $\Delta R =\allowbreak
\sqrt{(\Delta \phi)^2 + (\Delta\eta)^2}$ is the distance to the
nearest reconstructed track with momentum greater than
$250\,\mathrm{MeV}$ in the $\eta \-- \phi$ plane.  Isolation from
hadronic activity was imposed by requiring that the photon candidate
possessed at least $90 \%$ of the total energy of the jet-like object
of which it formed a part.  This also reduced the background of photon
candidates arising from neutral meson decay.  
 
Approximately 6000 events were selected at this stage; this sample was
dominated by background events in which one or more neutral mesons
such as $\pi^0$ and $\eta$, decaying to photons, produced a photon
candidate in the BEMC.

\section{Variables studied} 
In the previous ZEUS publication~\cite{pl:b715:88}, distributions of
photon and jet variables were studied. In the present analysis,
variables that depend on two of the three measured outgoing physical
objects were studied, namely the high-$p_T$ photon, the leading jet
and the scattered electron.  They were defined as follows:

\begin{itemize}
\item \xgamma\ is a measure of the fraction of the exchanged-photon
energy and longitudinal momentum that is given to the outgoing photon
and the jet:
$$\xgamma= \frac{E^\gamma - p_Z^\gamma + \Ejet - \pzjet}{2 E_{e}
\yJB},$$ where $E^\gamma$ and \Ejet\ denote the energies of the outgoing photon
and the jet, respectively, $p_Z^\gamma$ and \pzjet\ denote the
corresponding longitudinal momenta, $E_e=27.5$ GeV, and the
Jacquet--Blondel variable \yJB is given by
$\sum_\mathrm{EFO}(E^\mathrm{EFO} - p_{Z}^\mathrm{EFO})/2E_e$, summing
over all energy-flow objects in the event except the scattered
electron, each object being treated as equivalent to a massless particle.
This variable is sensitive to higher-order processes that generate additional 
particles in the event;
	
\item \xp\ estimates the fraction of the proton
energy transferred to the outgoing photon and jet:
$$\xp= \frac{E^\gamma + p_Z^\gamma + \Ejet\ + \pzjet}{2 E_{p}},$$
where $E_p = 920$ GeV.  This variable is sensitive to the partonic
structure of the proton;

\item $\Delta\phi $ is the azimuthal angle between the jet and the
outgoing photon: $\Delta\phi = |\phijet - \phi^\gamma|$, where
\phijet and $\phi^\gamma$ denote the azimuthal angles of the jet
and photon, respectively.  This variable is sensitive to the presence
of higher-order gluon radiation from the outgoing quark, which
generates a contribution to the non-collinearity between the photon
and the leading jet;

\item $\Delta\eta$ is the difference in pseudorapidity between the jet
and the outgoing photon: $\Delta\eta= \etajet - \eta^\gamma$, where \etajet\
and $\eta^\gamma$ denote the pseudorapidity of the jet and the photon,
respectively.  This variable is sensitive to the dynamical properties
of the scattering process;

\item \Dphiegamma\ is the azimuthal angle between the scattered
electron and the outgoing photon: $\Dphiegamma = |\phi^e -
\phi^\gamma|$, where $\phi^e$ denotes the azimuthal angle of the
electron; this and the following variable are sensitive to
higher-order processes and to whether the process is LL or QQ;

\item \Detaegamma\ is the difference in pseudorapidity between the
scattered electron and the photon: $\Detaegamma= \eta^e -
\eta^\gamma$, where $\eta^e$ denotes the pseudorapidity of the
electron.
\end{itemize}

A similar ZEUS analysis has been previously performed for
photoproduction~\cite{jhep:08:23}, studying all the present variables
except those associated with the scattered electron.

\section{Event simulation}
\label{sec-mc}
Monte Carlo (MC) event samples were generated to evaluate the detector
acceptance and to provide signal and background distributions. The
program \PYTHIA\ 6.416~\cite{jhep:0605:026} was used to simulate
prompt-photon emission for the study of the event-reconstruction
efficiency. In \PYTHIA, this process is simulated as a DIS process
with additional photon radiation from the quark line to account for QQ
photons.  Radiation from the lepton is not simulated.

The LL photons that were radiated into the detector and were isolated
from the outgoing electron were simulated using the generator {\sc
Djangoh} 6~\cite{djangoh}, an interface to the MC program {\sc
Heracles} 4.6.6~\cite{cpc:69:155}; higher-order QCD effects were
included using the colour dipole model of {\sc Ariadne}
4.12~\cite{cpc:71:15}.  Hadronisation of the partonic final state was
in each case performed by {\sc Jetset} 7.4~\cite{cpc:39:347} using the
Lund string model~\cite{pr:97:31}.  Interference between the LL and QQ
terms was neglected.

The main background to the QQ and LL photons came from photonic decays
of neutral mesons produced in general DIS processes. This background
was simulated using {\sc Djangoh} 6, within the same framework as the
LL events. This provided a realistic spectrum of single and multiple
mesons with well modelled kinematic distributions. 

The generated MC events were passed through ZEUS detector and trigger
 simulation programs based on {\sc Geant}
 3.21~\cite{tech:cern-dd-ee-84-1}. They were then reconstructed and
 analysed by the same programs as the data.

\section{Theoretical calculations}
\label{sec:theory}

The \PYTHIA\ predictions and the predictions of two parton-level
models were compared to the results of the present analysis.  The NLO
QCD calculation of Aurenche, Fontannaz and Guillet
(AFG)~\cite{epj:c44:395}, was performed in the
$\overline{\mathrm{MS}}$ scheme. Uncertainties on the QCD scale at
this order contribute a normalisation uncertainty of typically
$\pm8\%$.  This calculation was performed in the centre-of-mass frame
and transformed into the laboratory frame, which introduces
uncertainties on the cross sections in some regions of the parameter
space due to non-perturbative effects~\cite{af}.  The AFG predictions
were calculated with a cut of 2.5~GeV on the photon transverse 
momentum in the centre-of-mass frame, and do not include an LL
contribution, which was evaluated using the {\sc Djangoh--Heracles}
simulation and added separately to the AFG calculation for comparison
with the data.  The uncertainties on the AFG predictions shown in the
present paper represent the QCD scale uncertainties.

 A calculation by Baranov, Lipatov and Zotov
(BLZ)~\cite{pr:d81:094034} used updated parameters for the present
paper. It is based on the $k_T$-factorisation method. This approach
uses unintegrated parton densities and takes into account both QQ and
LL photons, neglecting the small interference contribution. The final
result is obtained as the convolution of the off-shell scattering
matrix element with the unintegrated quark distribution in the
proton. In the $k_T$-factorisation theory, some part of the
final-state jets can originate not only from the hard subprocess but
also from the parton evolution cascade in the initial state.  The
quoted uncertainties on the BLZ predictions represent the QCD
scale uncertainties.

In the previous ZEUS analysis of prompt photons in DIS, the measured
variables were associated with the entire event, with the outgoing
photon, and with jets.  Comparisons were made to an earlier NLO QCD
theory~\cite{np:b578:326,prl:96:132002,epj:c47:395} and to BLZ.  Both
theories described the shapes of the single-particle cross sections
well, but failed to reproduce the normalisation of the data. A later
version of the original AFG calculation agreed well with the
results~\cite{epj:c75:64}, and has been used in the present study.

The predictions of AFG and BLZ were calculated at the parton level and
incorporated kinematic and isolation criteria corresponding to the
data.  Corrections to the hadron level were made using \PYTHIA\ to
determine the ratio of the hadron-level cross sections to those at the
parton level for each variable in each bin.  The \PYTHIA\ events were
weighted at the parton level to represent the shapes of the AFG and
BLZ distributions in \xgamma\ in order to calculate the hadronisation
corrections for all the other measured variables.  The corrections for
AFG and BLZ were similar to within 10\%.  This procedure was also
applied separately to the AFG predictions for the different $Q^2$ ranges.

For the BLZ \xgamma\ distribution, 98\% of the parton-level cross
section is in the (0.9, 1.0) bin; consequently, for this variable a
transfer matrix from the parton to the hadron level was calculated
using \PYTHIA. The same procedure was used for the AFG \xgamma\
distribution.  The relevant transfer matrices for the other variables
gave similar results to the reweighting procedure.

\section{Extraction of the photon signal}
\label{sec:extraction}

The event sample selected according to the criteria described in
Section~\ref{sec-selec} was dominated by background from neutral meson
decays; thus the photon signal was extracted statistically following
the approach used in previous ZEUS analyses
\cite{pl:b413:201,*pl:b472:175,*pl:b511:19,pl:b595:86,pl:b687:16}.

The photon signal was evaluated making use of the width of
the BEMC energy-cluster corresponding to the photon candidate. This was
calculated as the variable $$\langle\delta Z\rangle=\sum \limits_i
E_i|Z_i-Z_{\mathrm{cluster}}| {\large/}( w_{\mathrm{cell}}\sum
\limits_i E_i),$$ where $Z_{i}$ is the $Z$ position of the centre of
the $i$-th cell, $Z_{\mathrm{cluster}}$ is the centroid of the EFO
cluster, $w_{\mathrm{cell}}$ is the width of the cell in the $Z$
direction, and $E_i$ is the energy recorded in the cell. The sum runs
over all BEMC cells in the EFO.

\begin{figure}[p]
\vfill
\begin{center}
\includegraphics[width=0.9\textwidth]{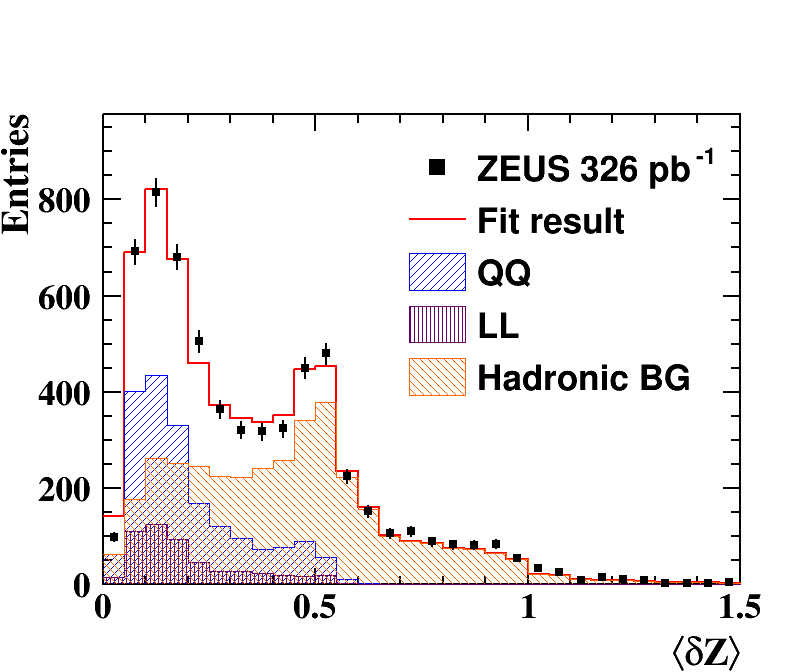}\\[-0.01\textwidth]
\end{center}
\caption{\small Distribution of $\langle \delta Z \rangle$ for the
full data sample.  The error bars represent the statistical
uncertainties on the data points.  The solid line shows a fit to the
data of three components with fixed shapes as described in the
text. The hatched histograms represent the LL and fitted  QQ components of the
fit and the fitted hadronic background (BG).  }
\label{fig:showers}
\vfill
\end{figure}

 The distributions of $\langle \delta Z \rangle$ for the full data set 
and the fitted MC are shown in Fig.\,\ref{fig:showers}.  The $\langle \delta
Z \rangle$ distribution exhibits a double-peaked structure with the
first peak at $\approx 0.1$, associated with the photon signal, and a
second peak at $\approx 0.5$, dominated by the $\pi^0\rightarrow\gamma
\gamma$ background. 

The contribution of isolated-photon events was determined for each bin
in each measured variable by a $\chi^2$ fit to the $\langle \delta Z
\rangle$ distribution in the range $0.05<\langle \delta Z \rangle <
0.8$, using the LL and QQ signal and background MC distributions as
described in Section~\ref{sec-mc}.  The mean value of $\chi^2$/n.d.f
was 1.2.  Compared to the earlier ZEUS publication~\cite{pl:b715:88},
improvements have been made in the modelling of the shapes of the
$\langle \delta Z \rangle$ distributions of the QQ and LL
contributions, using a comparison between the shapes associated with
the scattered electron in MC simulation of DIS and in real data.  By
treating the LL and QQ photons separately, account is taken of the
effect of their differing kinematic distributions on the acceptance,
and the effect of their differing ($\eta$, $E_{T}$) distributions on
the shape of the photon signal.

In performing the fit, the theoretically well determined LL
contribution was kept constant at its MC-predicted value and the other
components were varied.  Of the 6149 events selected, $2451\pm 102$
correspond to the extracted signal, including 526 LL photons. The
fitted scale factor applied to the QQ contribution in
Fig.~\ref{fig:showers} was 1.6, consistent with the earlier ZEUS
analysis.

For a given observable $Y$, the production cross section was
determined for each bin using
\begin{equation}
	\frac{d\sigma}{dY} = \frac{\mathcal{A}_{\mathrm{QQ}} 
        \cdot N(\gamma_{\mathrm{QQ}}) }{\mathcal{L} \cdot \Delta Y} + 
        \frac{d\sigma^{\mathrm{MC}}_{\mathrm{LL}}}{dY} \nonumber ,
\end{equation}
where $N(\gamma_{\mathrm{QQ}} )$ is the number of QQ photons extracted
from the fit, $\Delta Y$ is the bin width, $\mathcal{L}$ is the total
integrated luminosity, $\sigma^{\mathrm{MC}}_{\mathrm{LL}}$ is the
predicted cross section for LL photons from {\sc Djangoh--Hera\-cles} and
$\mathcal{A}_{\mathrm{QQ}}$ is the acceptance correction for QQ
photons.  The value of $\mathcal{A}_{\mathrm{QQ}}$ was calculated,
using the \PYTHIA\ MC, from the ratio of the number of events generated to
those reconstructed in a given bin; it lies in the range 0.91--2.28.
To improve the representation of the data, and hence the accuracy of
the acceptance corrections, the MC predictions were
reweighted.  This was done using parameterised functions of $Q^{2}$
and of $\eta^{\gamma}$, and also bin-by-bin as a function of photon energy;
the three reweighting factors were applied multiplicatively. Their
net effect on the acceptances was small.

\section{Systematic uncertainties}
The sources of systematic uncertainty on the measured cross sections
are as in the previous paper~\cite{pl:b715:88}.  The principal sources
of uncertainty were evaluated as follows:
\begin{itemize}
\item the energy scale of the photon candidate was varied by $\pm 2\%$. The
mean change of the cross section  was $\pm 6\%;$ 
\item the energy scale of the jets was varied by $\pm 1.5\%$ for jets
with $\ETjet > 10$ \GeV, $\pm 2.5\%$ for jets with \ETjet\ in the
range [6, 10] \GeV\ and $\pm 4\%$ for jets with $\ETjet < 6$ \GeV.  The
uncertainty was typically $\pm7\%$; 
\item the energy scale of the scattered electron was varied by $\pm2\%$.
The overall average effect on the cross sections was less than  $\pm 1\%$.
\end{itemize}

Systematic uncertainties related to the MC generators were evaluated
as follows:
\begin{itemize}
\item the dependence on the modelling of the hadronic background by means of
{\sc Djangoh--Hera\-cles} was investigated by varying the upper limit for the
$\langle \delta Z\rangle$ fit in the range $[0.6, 1.0]$, giving
variations that were typically $\pm5\%;$
\item uncertainties in the acceptance due to the \PYTHIA\ model were
accounted for by taking half of the change attributable to the
reweighting described in Section~\ref{sec:extraction} as a systematic
uncertainty; for most bins the effect was approximately $1 \%$.
\end{itemize}
\color{black}
Other sources of systematic uncertainty were found to be negligible
and were ignored~\cite{forrest:phd:2009,pl:b687:16}: these included
variations on the cuts on $\Delta R$, the track momentum, $E-p_Z$,
$Z_{\text{vtx}}$ and the electromagnetic fraction of the
photon shower, and a variation of 5\% on the LL fraction.  

The systematic uncertainties were symmetrised by taking the mean of
the positive and negative uncertainty values and were combined in
quadrature.  The common uncertainty of $1.8\% $ on the luminosity
measurement is not included in the tables and figures.

\section{Results}
\label{sec:results}
Differential cross sections for the production of an isolated photon
in DIS with an additional jet 
%, $ep\rightarrow e'\gamma+\mathrm{jet}$, 
have been measured in the laboratory frame in the kinematic region
defined by $4 < \ETgam< 15$~\gev, $-0.7<\etagam < 0.9$, $ \ETjet
>2.5$~\gev\ and $-1.5 <\etajet< 1.8$.  The DIS electron was
constrained to be in the angular range $\theta_e > 140^\circ,$ with
energy greater than 10 GeV and  $10 < Q^2 < 350$ \gev$^2,$ where
$Q^2$ was determined from the electron scattering angle.
The jets were formed according to the $k_T$-clustering algorithm with
the $R$ parameter set to 1.0. Photon isolation was imposed such that
at least $90 \%$ of the energy of the jet-like object containing the
photon belonged to the photon.

\begin{figure}[p]
\vfill
\begin{center}
\Large
\textbf{ZEUS }\\
\normalsize
\mbox{
\hspace*{-0.05\textwidth}
\includegraphics[width=0.52\textwidth]
{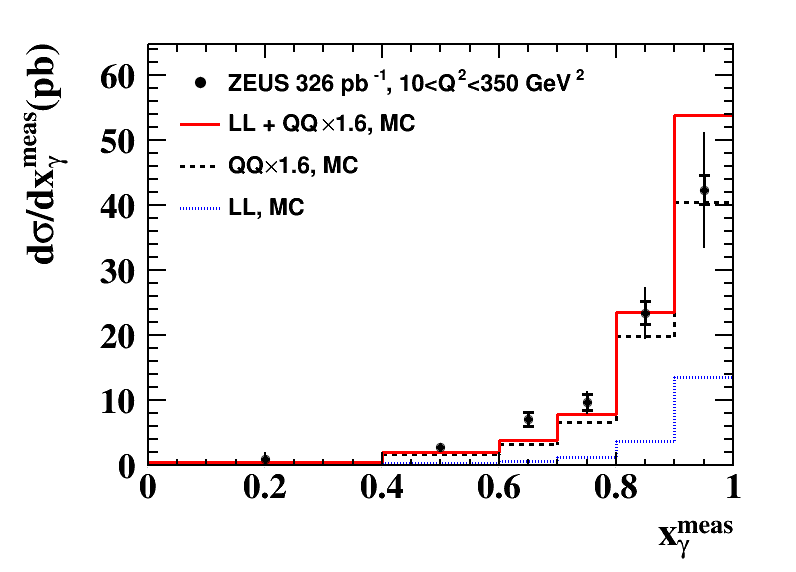}
\hspace*{-0.04\textwidth}
\includegraphics[width=0.52\textwidth]
{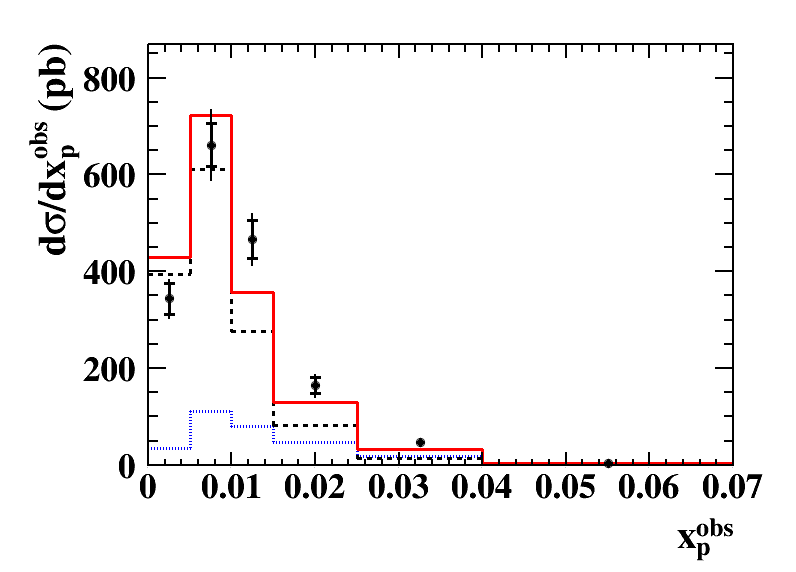}
}\\[-0.04\textwidth]
\hspace*{0.1\textwidth}(a)\hspace*{0.48\textwidth}(b)\\[0.04\textwidth]
\mbox{
\hspace*{-0.05\textwidth}
\includegraphics[width=0.52\textwidth]
{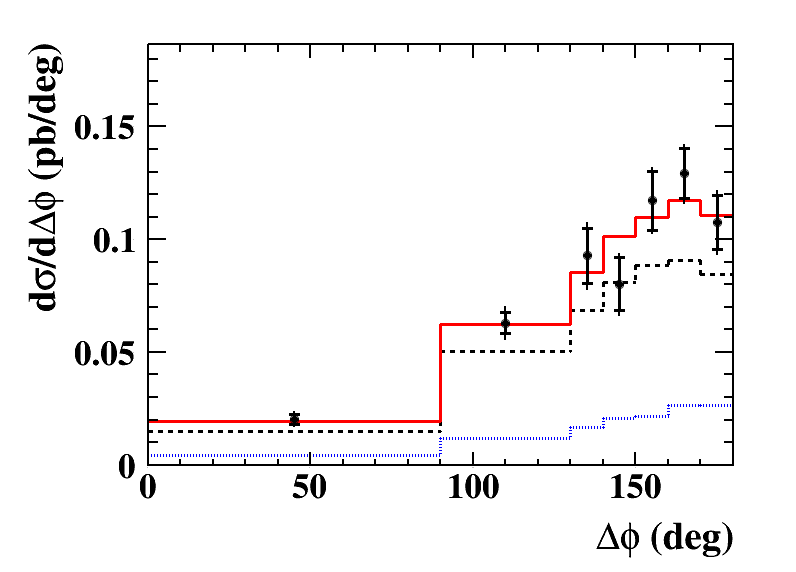}
\hspace*{-0.04\textwidth}
\includegraphics[width=0.52\textwidth]
{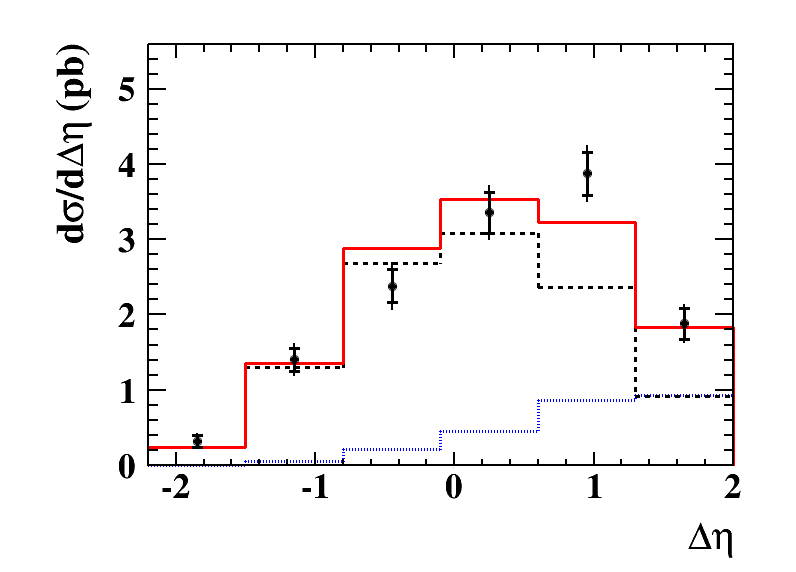}
}\\[-0.04\textwidth]
\hspace*{0.1\textwidth}(c)\hspace*{0.48\textwidth}(d)\\[0.04\textwidth]
\mbox{
\hspace*{-0.05\textwidth}
\includegraphics[width=0.52\textwidth]
{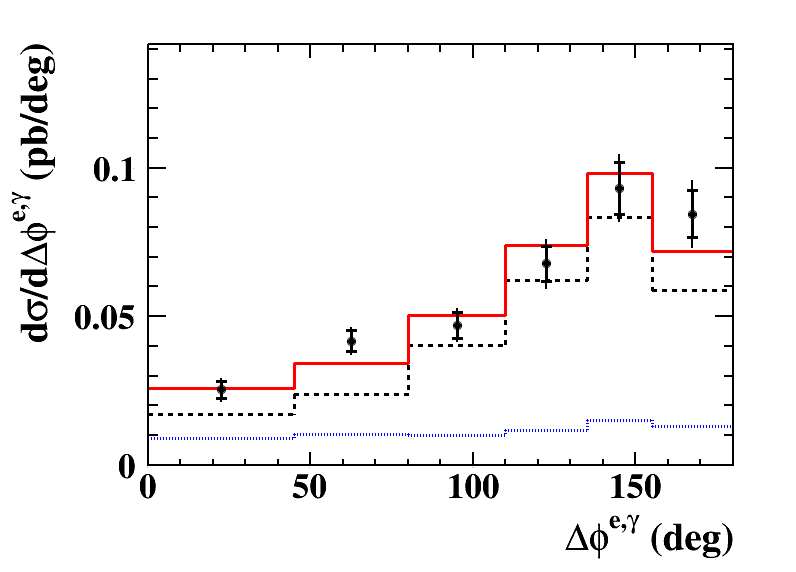}
\hspace*{-0.04\textwidth}
\includegraphics[width=0.52\textwidth]
{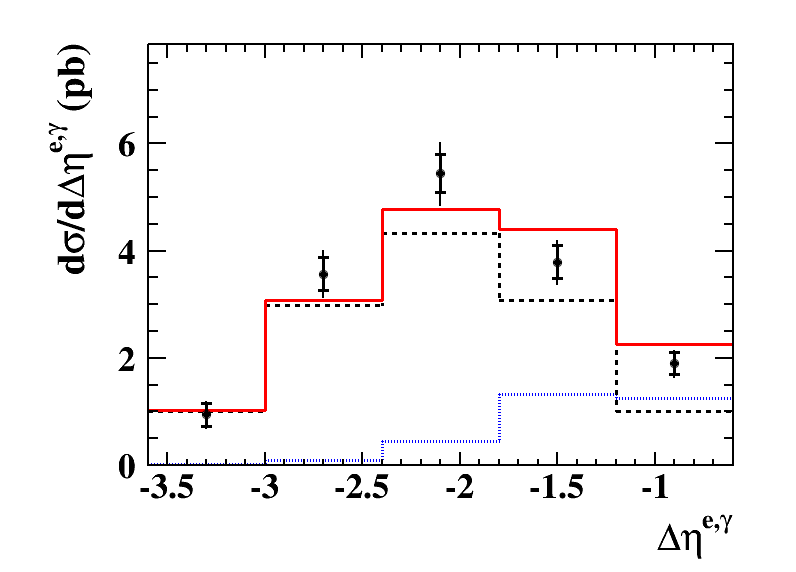}
}
\\[-0.04\textwidth]
\hspace*{0.1\textwidth}(e)\hspace*{0.48\textwidth}(f)\\[0.04\textwidth]
\end{center}
\caption{\small Differential cross sections in (a) \xgamma, (b)
\xp, (c) $\Delta\phi $, (d) $\Delta\eta$, (e)
\Dphiegamma, and (f) \Detaegamma, 
for the full range  $10<Q^2 < 350$ GeV\/$^2$.
The inner and outer error bars show, respectively, the statistical
uncertainty and the statistical and systematic uncertainties added in
quadrature. The solid histograms are the Monte Carlo
predictions from the sum of QQ photons from \PYTHIA\ normalised by
a factor 1.6 plus {\sc Djangoh--Heracles} LL photons. The dashed (dotted) lines
show the QQ (LL) contributions.}
\label{fig:xsec1}
\vfill
\end{figure}

The differential cross sections for the full $Q^2$ range as functions
of $ \xgamma,\;\xp,\; \Delta\phi,\; \Delta\eta,$ \Dphiegamma and
\Detaegamma\ are shown in Fig.~\ref{fig:xsec1} and are given in Tables
\ref{tab:dsdxgamma}--\ref{tab:dsddeta_e_ph}, which also list the
values of the LL contributions and the hadronisation corrections.  The
cross section decreases with increasing \xp, having a peak around
0.01, and rises at high values of $\xgamma, \Delta\phi$ and
\Dphiegamma.  The predictions for the sum of the expected LL
contribution from {\sc Djangoh--Heracles} and a factor of 1.6 times
the expected QQ contribution from \PYTHIA\ agree well with the
measurements.  The success of the \PYTHIA\ calculation can be
attributed to its use of a leading-logarithm approach to gluon
emission to augment its LO parton-scattering calculation.

\begin{figure}[p]
\vfill
\begin{center}
\Large
\textbf{ZEUS }\\
\normalsize
\mbox{
\hspace*{-0.05\textwidth}
\includegraphics[width=0.52\textwidth]{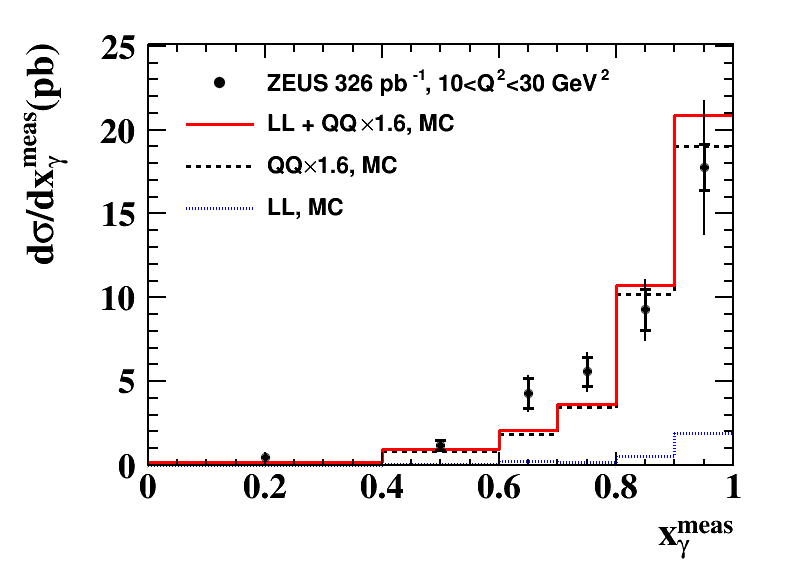}
\hspace*{-0.04\textwidth}
\includegraphics[width=0.52\textwidth]{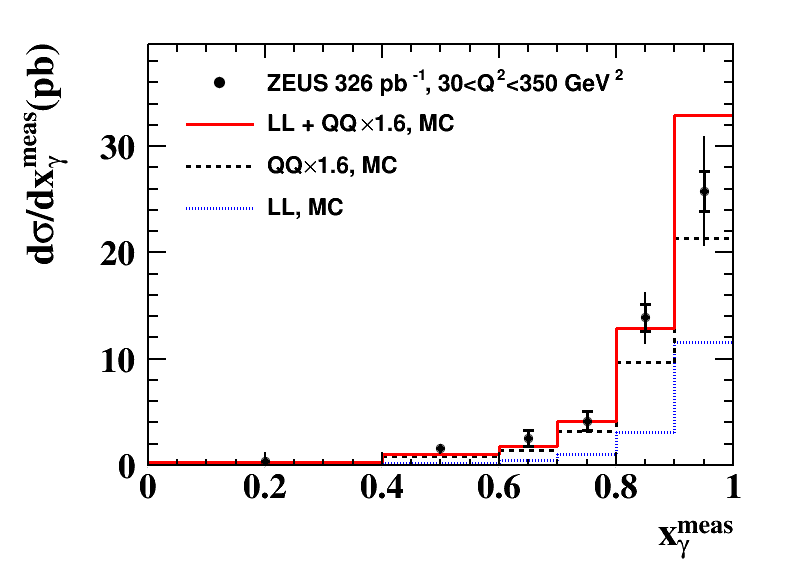}
} \\[-0.04\textwidth]
\hspace*{0.1\textwidth}(a)\hspace*{0.48\textwidth}(b)\\[0.04\textwidth]
\mbox{
\hspace*{-0.05\textwidth}
\includegraphics[width=0.52\textwidth]{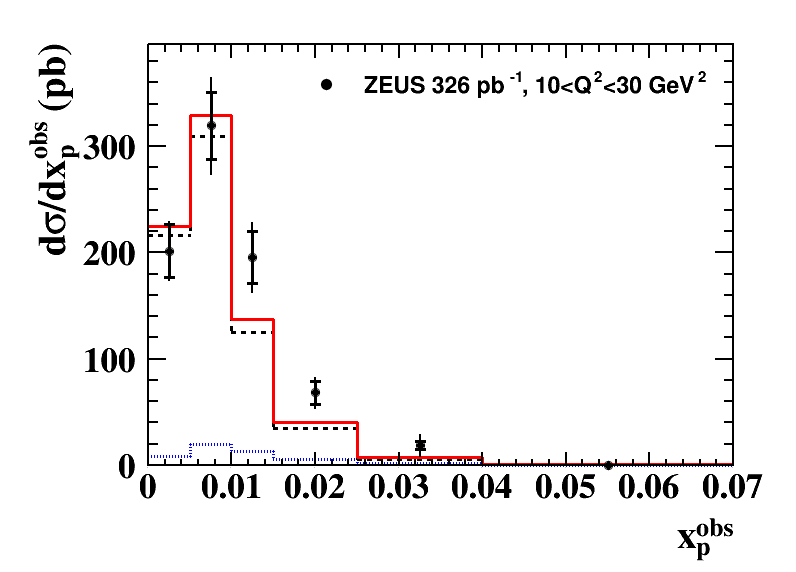}
\hspace*{-0.04\textwidth}
\includegraphics[width=0.52\textwidth]{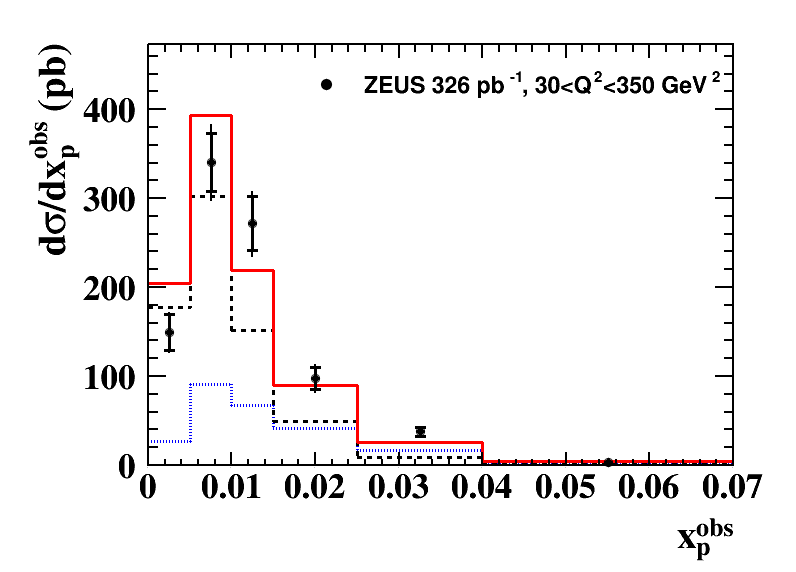}
}\\[-0.04\textwidth]
\hspace*{0.1\textwidth}(c)\hspace*{0.48\textwidth}(d)\\[0.04\textwidth]
\mbox{
\hspace*{-0.05\textwidth}
\includegraphics[width=0.52\textwidth]{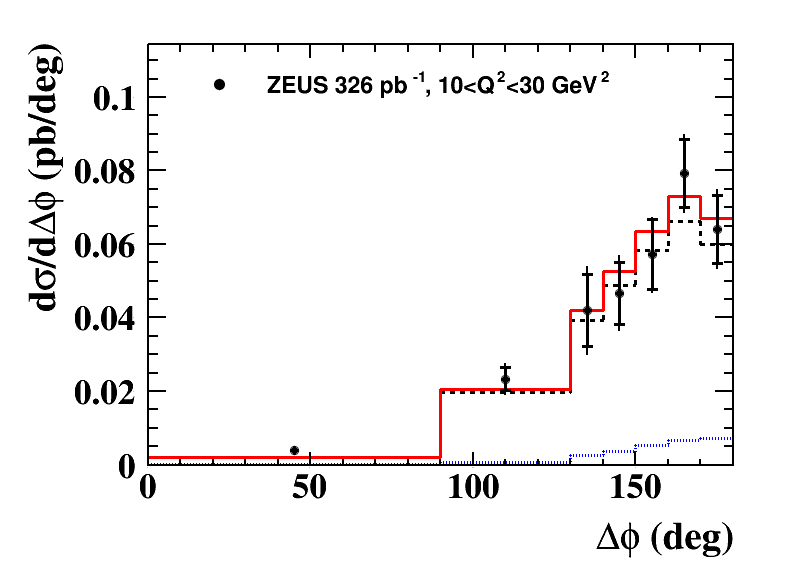}
\hspace*{-0.04\textwidth}
\includegraphics[width=0.52\textwidth]{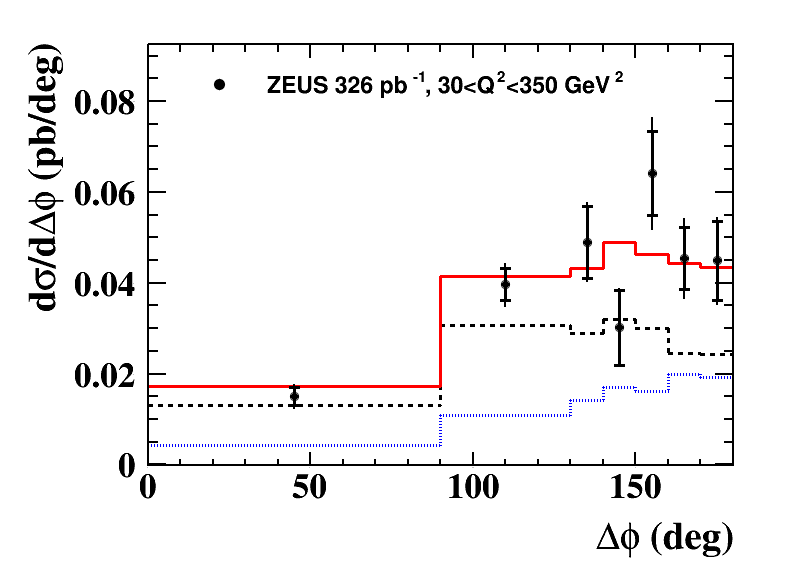}
}
\\[-0.04\textwidth]
\hspace*{0.1\textwidth}(e)\hspace*{0.48\textwidth}(f)\\[0.04\textwidth]
\end{center}
\caption{\small Differential cross sections 
for the regions $10<Q^2 < 30$ and $30<Q^2 <350$ GeV\/$^2$:
(a, b) \xgamma, (c, d) \xp, and (e, f) $\Delta\phi \,\,$.
The inner and outer error bars show, respectively, the statistical
uncertainty and the statistical and systematic uncertainties added in
quadrature. The solid histograms are the Monte Carlo
predictions from the sum of QQ photons from \PYTHIA\ normalised by
a factor 1.6 plus {\sc Djangoh--Heracles} LL photons. The dashed (dotted) lines
show the QQ (LL) contributions.}
\label{fig:xsec4}
\vfill
\end{figure}

\begin{figure}[p]
\vfill
\begin{center}
\Large
\textbf{ZEUS }\\
\normalsize
\mbox{
\hspace*{-0.05\textwidth}
\includegraphics[width=0.52\textwidth]{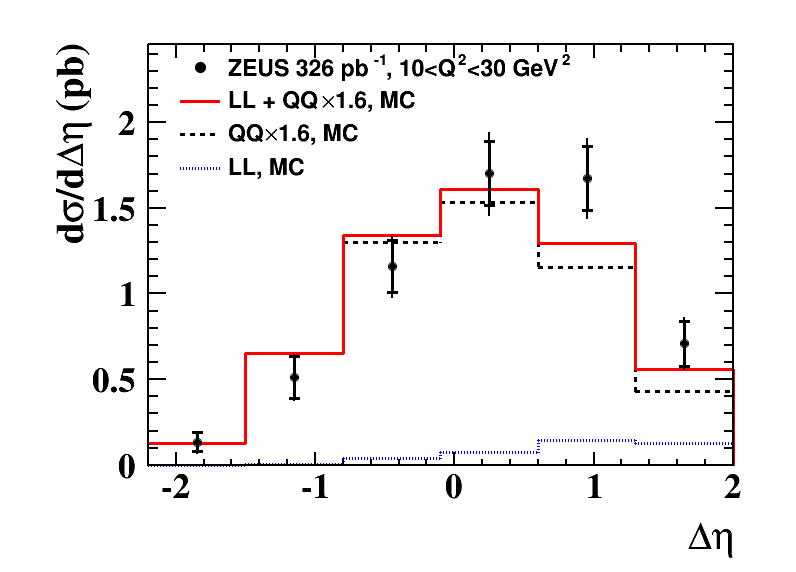}
\hspace*{-0.04\textwidth}
\includegraphics[width=0.52\textwidth]{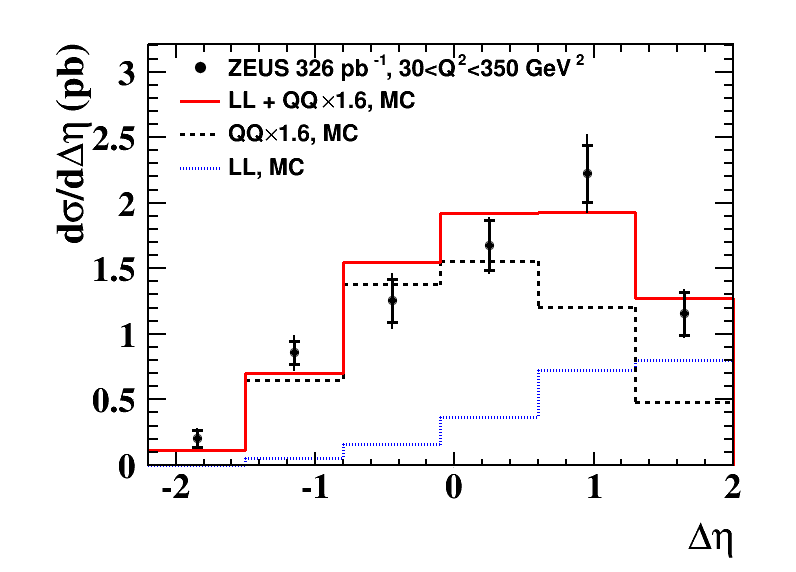}
} \\[-0.04\textwidth]
\hspace*{0.1\textwidth}(a)\hspace*{0.48\textwidth}(b)\\[0.04\textwidth]
\mbox{
\hspace*{-0.05\textwidth}
\includegraphics[width=0.52\textwidth]{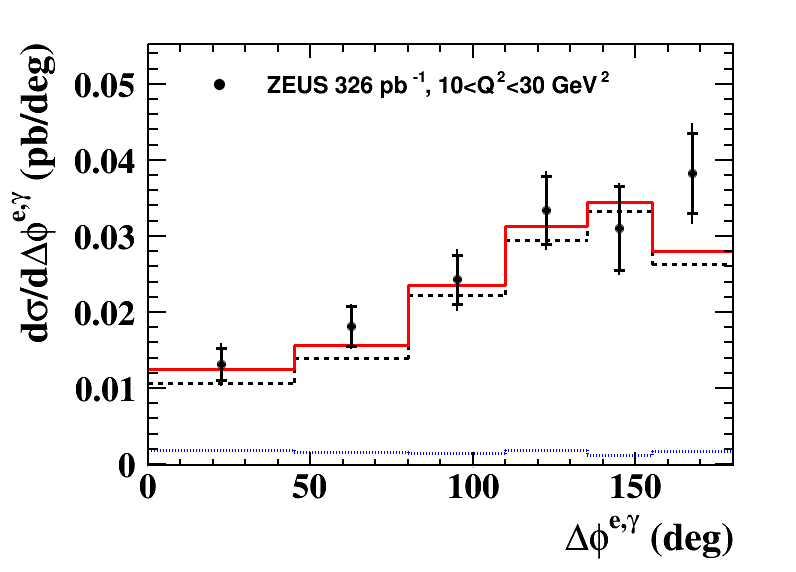}
\hspace*{-0.04\textwidth}
\includegraphics[width=0.52\textwidth]{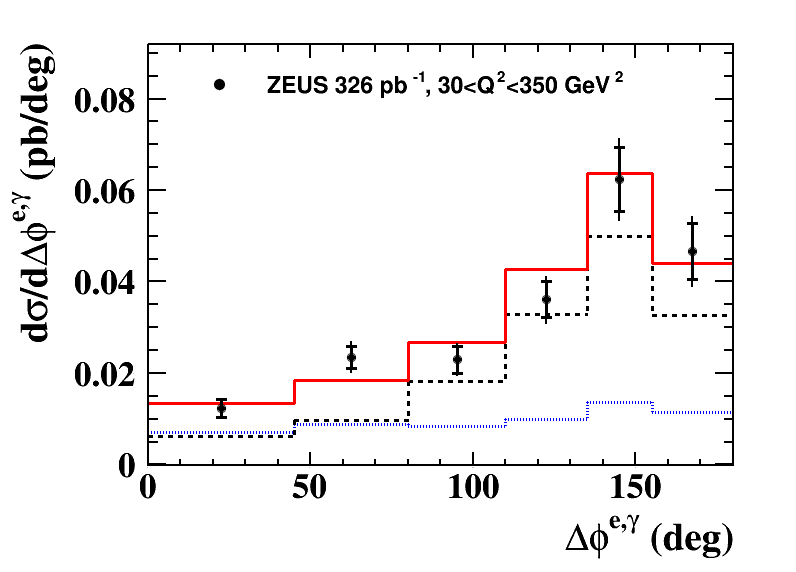}
}\\[-0.04\textwidth]
\hspace*{0.1\textwidth}(c)\hspace*{0.48\textwidth}(d)\\[0.04\textwidth]
\mbox{
\hspace*{-0.05\textwidth}
\includegraphics[width=0.52\textwidth]{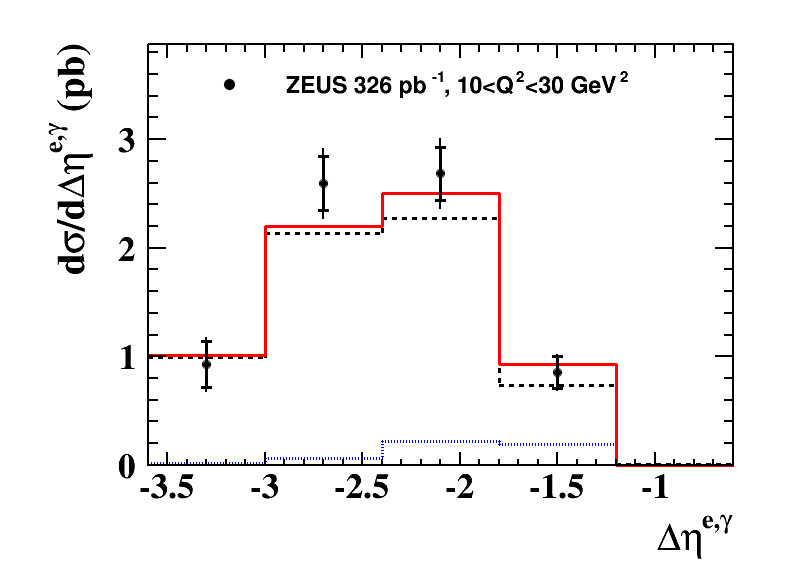}
\hspace*{-0.04\textwidth}
\includegraphics[width=0.52\textwidth]{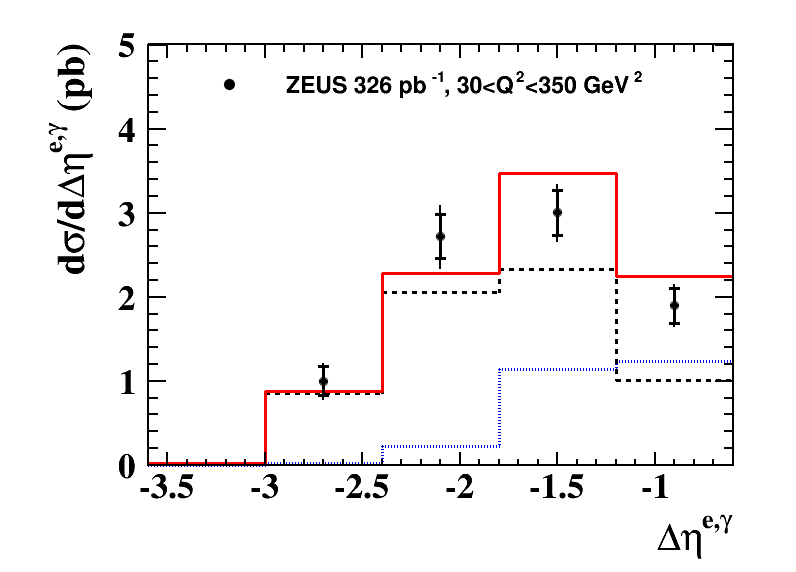}
}
\\[-0.04\textwidth]
\hspace*{0.1\textwidth}(e)\hspace*{0.48\textwidth}(f)\\[0.04\textwidth]
\end{center}
\caption{\small Differential cross sections 
for the regions $10<Q^2 < 30$ and $30<Q^2 <350$ GeV$^2$:
(a, b) $\Delta\eta$, 
(c, d) \Dphiegamma, and (e, f) \Detaegamma.
The inner and outer error bars show, respectively, the statistical
uncertainty and the statistical and systematic uncertainties added in
quadrature. The solid histograms are the Monte Carlo
predictions from the sum of QQ photons from \PYTHIA\ normalised by
a factor 1.6 plus {\sc Djangoh--Heracles} LL photons. The dashed (dotted) lines
show the QQ (LL) contributions.}
\vfill
\label{fig:xsec5}
\end{figure}

The differential cross sections for the separate ranges
$10<Q^2<30$~\gev$^2$ and $30 <Q^2<350$~\gev$^2$ are shown in
Figs.~\ref{fig:xsec4} and \ref{fig:xsec5}.  In both these ranges, a
good description of the data is given by the combination of the LL and
\PYTHIA\ MCs.  The LL contribution is small in the lower $Q^2$ region,
as was already seen in Fig.\ 3(a) of the earlier ZEUS
publication~\cite{pl:b715:88}.  In the higher $Q^2$ range, the LL
component contributes significantly, as can be seen in the \xp,
$\Delta\phi,\;\Delta\eta,$ and \Detaegamma distributions where it is
dominant at high values of these variables.  This reflects the changes
with $Q^2$ in the structure of the contributing processes.

\begin{figure}[p]
\vfill
\begin{center}
\Large
\textbf{ZEUS }\\
\normalsize
\mbox{
\hspace*{-0.05\textwidth}
\includegraphics[width=0.52\textwidth]{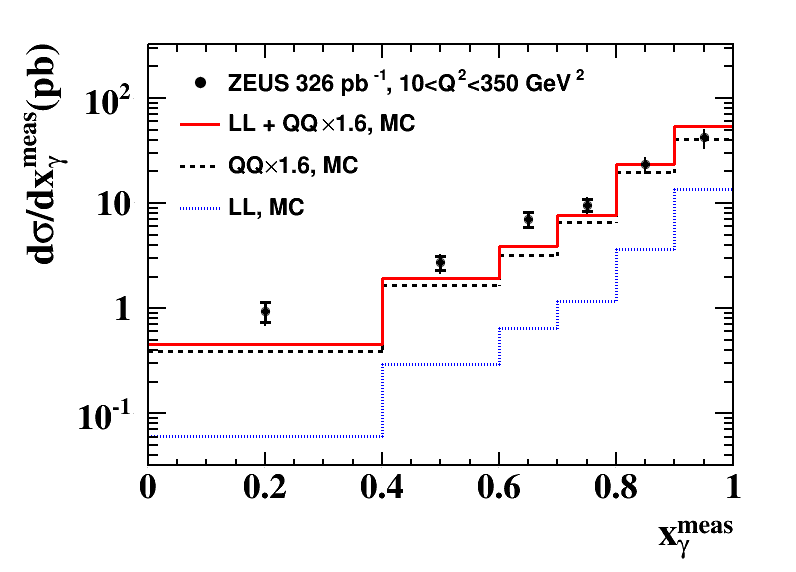}
\hspace*{-0.04\textwidth}
\includegraphics[width=0.52\textwidth]{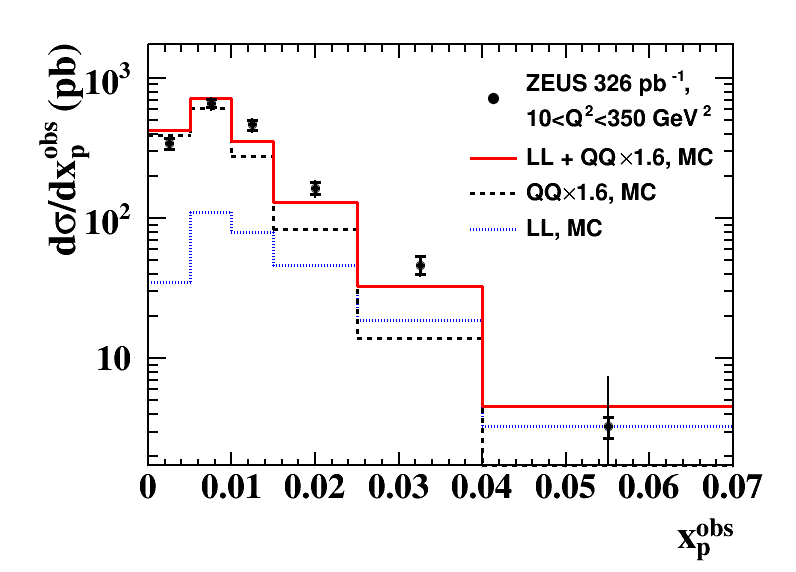}
} \\[-0.04\textwidth]
\hspace*{0.1\textwidth}(a)\hspace*{0.48\textwidth}(b)\\[0.04\textwidth]
\mbox{
\hspace*{-0.05\textwidth}
\includegraphics[width=0.52\textwidth]{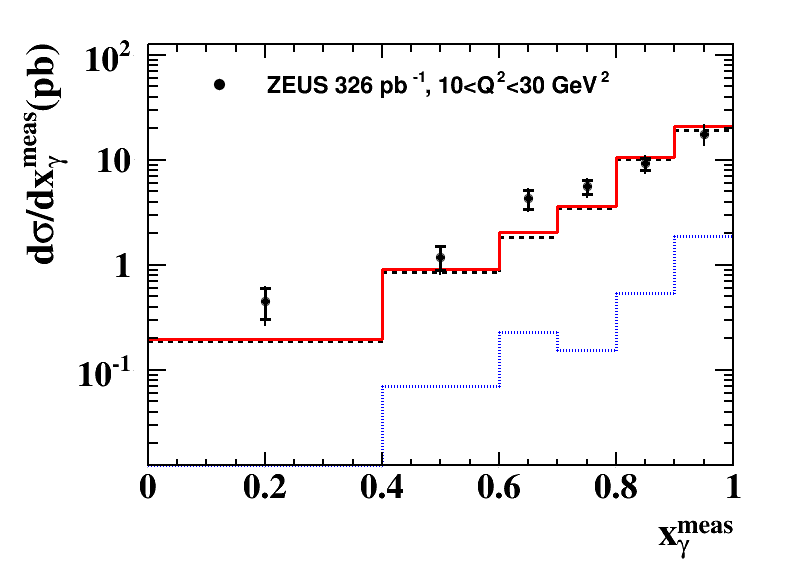}
\hspace*{-0.04\textwidth}
\includegraphics[width=0.52\textwidth]{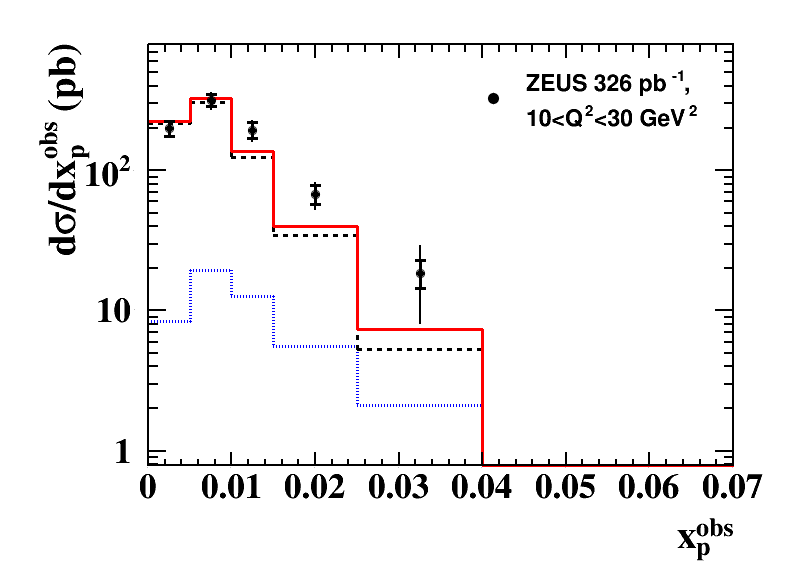}
}\\[-0.04\textwidth]
\hspace*{0.1\textwidth}(c)\hspace*{0.48\textwidth}(d)\\[0.04\textwidth]
\mbox{
\hspace*{-0.05\textwidth}
\includegraphics[width=0.52\textwidth]{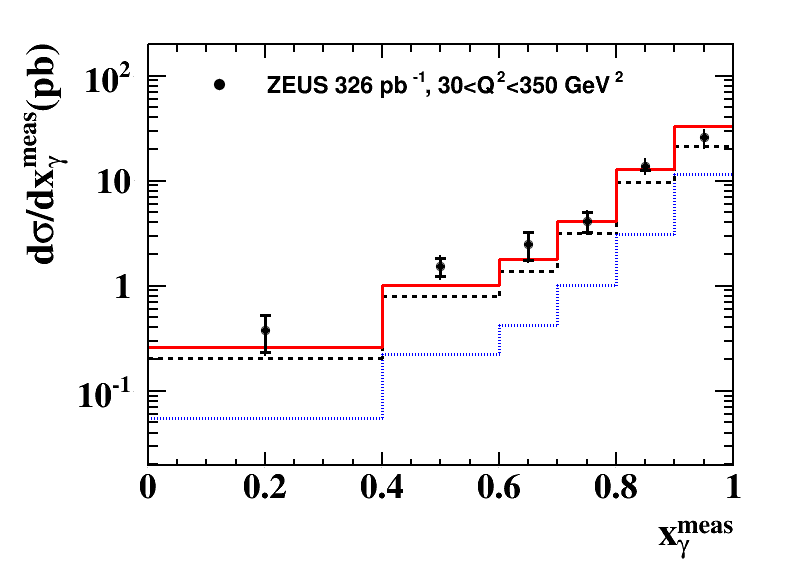}
\hspace*{-0.04\textwidth}
\includegraphics[width=0.52\textwidth]{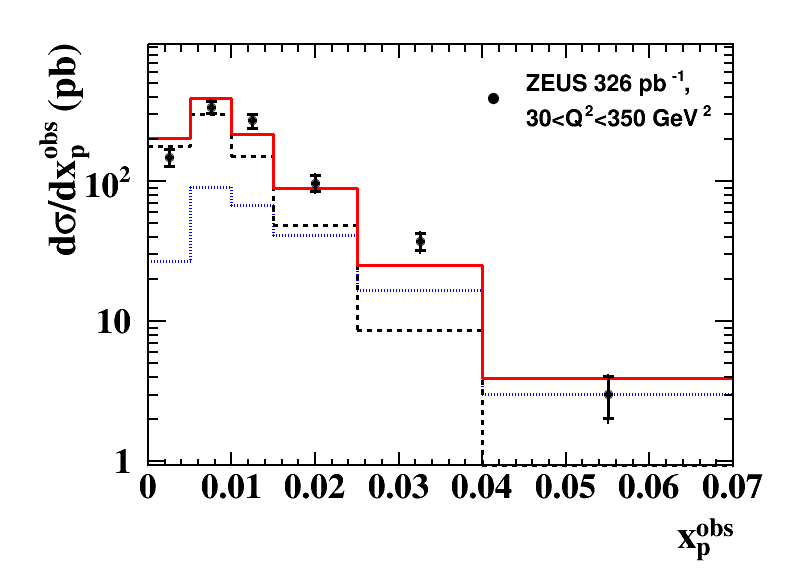}
}
\\[-0.04\textwidth]
\hspace*{0.1\textwidth}(e)\hspace*{0.48\textwidth}(f)\\[0.04\textwidth]
\end{center}
\caption{\small Differential cross sections in (a, c, e) \xgamma\ and
(b, d, f) \xp\ for (a, b) $10<Q^2 < 350$~\gev$^2$, (c, d) $10<Q^2
<30$~\gev$^2$, and (e, f) $30<Q^2 <350$~\gev$^2$. The distributions
are as shown in Figs.~\ref{fig:xsec1} -- \ref{fig:xsec5} but with
logarithmic vertical scale.}
\label{fig:logfig}
\vfill
\end{figure}

The increased importance of the LL component at higher $Q^2$ is also
reflected in the \xgamma\ distribution.  Figure \ref{fig:logfig}
presents the \xgamma\ and \xp\ cross sections on a logarithmic scale.
The data in the low-\xgamma\ region are satisfactorily described by \PYTHIA\ 
without the need for further higher-order processes.

Comparisons of the data with the AFG and BLZ predictions are presented
for the entire $Q^2$ range in Fig.~\ref{fig:xsec6}.  The updated BLZ
predictions describe the shape of most of the distributions reasonably
well, but there is an overestimation of about 20\% in the overall
cross section, and the extremely peaked prediction for the
\xgamma\ distribution is not in agreement with the data.  The AFG
predictions describe all the distributions well and also agree in the
overall normalisation.

Comparisons of the data with the AFG model in the two separate $Q^2$
ranges are shown in Figs.~\ref{fig:xsec7}--\ref{fig:xsec8}.  In the
higher $Q^2$ range, the description by AFG is excellent.  In the lower
range, the only deviation observable is in the $\Delta\eta$
distribution, where the data show a tendency towards higher values
than the theory.  This might be related to the cut of 2.5~GeV on the
transverse photon momentum applied in the AFG
calculation~\cite{epj:c44:395}.

\section{Summary}

The production of isolated photons accompanied by jets has been
measured in deep inelastic scattering with the ZEUS detector at HERA,
using an integrated luminosity of $326\,\mathrm{pb}^{-1}$. Expanding
on earlier ZEUS results~\cite{pl:b715:88}, which studied
single-particle distributions, differential cross sections have been
evaluated as functions of pairs of measured variables in combination.
The kinematic region in the laboratory frame was defined by $4 <
\ETgam< 15\,\mathrm{\gev} $, $-0.7<\etagam < 0.9$, $\ETjet >2.5$~\gev\
and $-1.5 <\etajet< 1.8$.  The DIS electron was constrained to be in
the angular range $\theta_e > 140^\circ,$ with energy greater than 10
GeV and $10 < Q^2 < 350$ \gev$^2,$ where $Q^2$ was determined from the
electron scattering angle.  The jets were formed according to the
$k_T$-clustering algorithm with the $R$ parameter set to 1.0. Photon
isolation was imposed such that at least $90 \%$ of the energy of the
jet-like object containing the photon belonged to the photon.
Differential cross sections are presented for the following variables:
the fraction of the incoming photon energy and momentum that is
transferred to the outgoing photon and the leading jet; the fraction
of the incoming proton energy transferred to the photon and leading
jet; the differences in azimuthal angle and pseudorapidity between the
outgoing photon and the leading jet and between the outgoing photon
and the scattered electron.

The \PYTHIA\ prediction for the quark-radiated photon component plus
the {\sc Djangoh--Heracles} calculation for the lepton-radiated
component describes all the distributions well if the \PYTHIA\ prediction 
is scaled up by a factor of 1.6.  This is also true if the
data are divided into ranges above and below a value of $Q^2 = 30$
GeV$^2$.  Predictions from two theoretical models were also compared
to the data.  The BLZ model gives a fair description of the data but
does not give a good description of the overall normalisation or the
shape of some of the distributions.  The AFG model gives an excellent
description of the normalisation and almost all the distributions,
both for the entire data set and for the separate $Q^2$ ranges.

% ----------------------------------------------------------------------------
%       Mandatory acknowledgements. You may add your buddies to it.
% ----------------------------------------------------------------------------
\section*{Acknowledgements}
\label{sec-ack}

\Zacknowledge\ We also thank P.~Aurenche, M.~Fontannaz and A.~Lipatov
for providing theoretical results and express our appreciation for the
contributions from our much-missed late colleague, Nikolai Zotov.

\vfill\eject

%------------------------------------------------------------------------------
%       Bibliography
%------------------------------------------------------------------------------
\providecommand{\etal}{et al.\xspace}
\providecommand{\coll}{Collaboration}
\catcode`\@=11
\def\@bibitem#1{%
\ifmc@bstsupport
  \mc@iftail{#1}%
    {;\newline\ignorespaces}%
    {\ifmc@first\else.\fi\orig@bibitem{#1}}
  \mc@firstfalse
\else
  \mc@iftail{#1}%
    {\ignorespaces}%
    {\orig@bibitem{#1}}%
\fi}%
\catcode`\@=12

\clearpage

%------------------------------------------------------------------------------
%       Tables
%------------------------------------------------------------------------------
\newcommand{\TO}{\hspace*{-1.5ex}--\hspace*{-1.5ex}}
\newcommand{\BR}{\hspace*{1.2ex}}
\newcommand{\B}{$\hspace*{1.2ex}$}
\newcommand{\sta}{\mathrm{(stat.)}}
\newcommand{\sys}{\mathrm{(sys.)}}
\renewcommand{\strut}{\rule[-1.4ex]{0ex}{4ex}}
\newcommand{\strutb}{\rule[-0.4ex]{0ex}{3ex}}
%-------------------------------------------------------------------------------
%       Results
%------------------------------------------------------------------------------
\newpage

%%%%%%%%%%%%%%%%%%%%%
\newcommand{\dsigLL}{\ensuremath{d\sigma_{\mathrm{LL}}}}
% Table 1===========
\begin{table}
\centering
%\begin{adjustbox}{width=1\textwidth}
\begin{tabular}{|l @{ -- } r | 
c @{${}\pm{}$} 
c @{(stat.)${}\pm{}$} 
c @{(sys.)\hspace{\tabcolsep}} |
c @{${}\pm{}$} 
c @{(stat.)\hspace{\tabcolsep}} |
c @{\hspace{\tabcolsep}} 
|}
\hline
\multicolumn{2}{|c|}{\begin{tabular}{c}\strutb$\xgamma$ \\range\\ \end{tabular}} 
& \multicolumn{3}{ c|}{ {\large $\frac{d\sigma}{d\xgamma}$   ($\mathrm{pb}$)}} 
& \multicolumn{2}{ c|}{ {\large $\frac{\dsigLL}{d\xgamma}$ ($\mathrm{pb}$)}} 
& \multicolumn{1}{c|}{\begin{tabular}{c}had. \\ cor.\\ \end{tabular}} \\
\hline
\hline
\multicolumn{8}{|c|}{\strut $10 < Q^2 < 350$ GeV$^2$ }            \\
\hline
0.0 & 0.4 &  0.94 & 0.20 &   0.11 &  0.06 &   0.01 & 0.63 \\
0.4 & 0.6 &  2.73 & 0.43 &   0.32 &  0.29 &   0.04 & 0.90 \\
0.6 & 0.7 &  7.06 & 1.14 &   0.38 &  0.65 &   0.09 & 1.27 \\
0.7 & 0.8 &  9.64 & 1.24 &   1.06 &  1.17 &   0.12 & 1.93 \\
0.8 & 0.9 & 23.40 & 1.75 &   3.51 &  3.67 &   0.22 & 2.06 \\
0.9 & 1.0 & 42.34 & 2.26 &   8.54 & 13.49 &   0.42 & 0.64 \\
\hline
\multicolumn{8}{|c|}{\strut $10 < Q^2 < 30$ GeV$^2$ }             \\
\hline
0.0& 0.4 &     0.45 & 0.15 &   0.09 &  0.01 &   0.01 & 0.68  \\
0.4& 0.6 &     1.19 & 0.31 &   0.18 &  0.07 &   0.02 & 1.00  \\
0.6& 0.7 &     4.30 & 0.88 &   0.49 &  0.23 &   0.06 & 1.30  \\
0.7& 0.8 &     5.58 & 0.88 &   0.69 &  0.16 &   0.04 & 2.02  \\
0.8& 0.9 &     9.27 & 1.20 &   1.32 &  0.54 &   0.08 & 2.11  \\
0.9& 1.0 &    17.76 & 1.37 &   3.73 &  1.89 &   0.16 & 0.63  \\
\hline
\multicolumn{8}{|c|}{\strut $30 \leq Q^2 < 350$ GeV$^2$ }  \\
\hline
0.0& 0.4 &     0.38 & 0.15 &   0.05 &  0.06 &   0.01 & 0.60  \\
0.4& 0.6 &     1.55 & 0.30 &   0.23 &  0.22 &   0.04 & 0.82  \\
0.6& 0.7 &     2.50 & 0.73 &   0.36 &  0.42 &   0.07 & 1.25  \\
0.7& 0.8 &     4.15 & 0.89 &   0.53 &  1.01 &   0.11 & 1.86  \\
0.8& 0.9 &    13.90 & 1.27 &   2.01 &  3.14 &   0.20 & 2.02  \\
0.9& 1.0 &    25.81 & 1.89 &   4.74 & 11.61 &   0.38 & 0.65  \\
\hline
\end{tabular}
%\end{adjustbox}
\caption{Measured differential cross-section
$\frac{d\sigma}{d\xgamma}$. The quoted systematic uncertainty includes
all the components added in quadrature.  The calculated LL
contribution which was added to the \PYTHIA\ and AFG calculations is
also listed, and the hadronisation correction calculated for the AFG
predictions.  Differences between cross sections in the first section
and the sum of the corresponding values in the second and third
sections are of statistical origin.
\label{tab:dsdxgamma}}
\end{table}

% Table 2===========
\begin{table}
\centering
%\begin{adjustbox}{width=1\textwidth}
\begin{tabular}{|l @{ -- } r | 
c @{${}\pm{}$} 
c @{(stat.)${}\pm{}$} 
c @{(sys.)\hspace{\tabcolsep}} |
c @{${}\pm{}$} 
c @{(stat.)\hspace{\tabcolsep}} |
c @{\hspace{\tabcolsep}} 
|}
\hline
\multicolumn{2}{|c|}{\begin{tabular}{c}\strutb\xp\ \\range\\ \end{tabular}} 
& \multicolumn{3}{ c|}{ {\large $\frac{d\sigma}{d\xp}$   ($\mathrm{pb}$)}} 
& \multicolumn{2}{ c|}{ {\large $\frac{\dsigLL}{d\xp}$ ($\mathrm{pb}$)}} 
& \multicolumn{1}{c|}{\begin{tabular}{c}had. \\ cor.\\ \end{tabular}} \\

\hline
\hline
\multicolumn{8}{|c|}{\strut $10 < Q^2 < 350$ GeV$^2$ }  \\
\hline
0.000 & 0.005 &   344.3 &   31.7 &  22.9 & 35.2 &   3.0 & 0.69 \\
0.005 & 0.010 &   661.8 &   45.3 &  56.6 & 110.8 &   5.3 & 0.81 \\
0.010 & 0.015 &   467.1 &   38.9 &  35.5 & 80.0 &   4.5 & 0.91 \\
0.015 & 0.025 &   164.5 &   16.5 &  16.1 & 46.6 &   2.4 & 0.99 \\
0.025 & 0.040 &    46.7 &    6.8 &   2.7 & 18.7 &   1.3 & 1.06 \\
0.040 & 0.070 &     3.3 &    0.6 &   2.1 &  3.3 &   0.4 & 1.00 \\
\hline
\multicolumn{8}{|c|}{\strut $10 < Q^2 < 30$ GeV$^2$ }  \\
\hline
0.000 & 0.005 &   201.8 &   25.0 &  11.1 &  8.4 &   1.4 & 0.71  \\
0.005 & 0.010 &   319.6 &   31.4 &  31.8 & 19.4 &   2.2 & 0.84  \\
0.010 & 0.015 &   195.5 &   24.5 &  20.5 & 12.7 &   1.8 & 0.98  \\
0.015 & 0.025 &    68.1 &   10.4 &   9.8 &  5.6 &   0.9 & 1.03  \\
0.025 & 0.040 &    18.7 &    4.1 &   9.5 &  2.1 &   0.4 & 1.08  \\
0.040 & 0.070 &    0.2  &   0.1  &  0.1  & 0.2  &   0.1  & 0.95  \\
\hline
\multicolumn{8}{|c|}{\strut $30 \leq Q^2 < 350$ GeV$^2$ }   \\
\hline

0.000 & 0.005 &   149.3 &   20.0 &   9.1 & 26.8   &  2.6    & 0.68  \\
0.005 & 0.010 &   340.7 &   32.9 &  25.0 & 91.4   &  4.8    & 0.78  \\
0.010 & 0.015 &   271.7 &   30.5 &  17.4 & 67.3   &  4.1    & 0.88  \\
0.015 & 0.025 &    97.7 &   12.8 &   8.1 & 41.0   &  2.3    & 0.97  \\
0.025 & 0.040 &    37.5 &    5.3 &   3.1 & 16.6   &  1.2    & 1.06  \\
0.040 & 0.070 &     3.0 &    1.0 &   2.1 &  3.0   &  0.4    & 1.01  \\
\hline
\end{tabular}
%\end{adjustbox}
\caption{Measured differential cross-section $\frac{d\sigma}{d\xp}$. Details as in Table \ref{tab:dsdxgamma}. \label{tab:dsdxp}}
\end{table}

% Table 3===========
\begin{table}
\centering
%\begin{adjustbox}{width=1\textwidth}
\begin{tabular}{|l @{ -- } r | 
c @{${}\pm{}$} 
c @{(stat.)${}\pm{}$} 
c @{(sys.)\hspace{\tabcolsep}} |
c @{${}\pm{}$} 
c @{(stat.)\hspace{\tabcolsep}} |
c @{\hspace{\tabcolsep}} 
|}
\hline
\multicolumn{2}{|c|}{\begin{tabular}{c}\strutb$\Delta\phi$ \\ range \\ (deg)\\ \end{tabular}} 
& \multicolumn{3}{ c|}{ {\large $\frac{d\sigma}{d\Delta\phi}$   ($\mathrm{pb/deg}$)}} 
& \multicolumn{2}{ c|}{ {\large $\frac{\dsigLL}{d\Delta\phi}$ ($\mathrm{pb/deg}$)}} 
& \multicolumn{1}{c|}{\begin{tabular}{c}had. \\ cor.\\ \end{tabular}} \\

\hline\hline
\multicolumn{8}{|c|}{\strut $10 < Q^2 < 350$ GeV$^2$ }  \\
\hline
0 & 90    & 0.020  & 0.002 & 0.003  &  0.004 &  0.001 & 0.68  \\
90 & 130  & 0.063 & 0.005 & 0.005 &  0.012 &  0.001 & 0.82  \\
130 & 140 & 0.093 & 0.012 & 0.008 &  0.017 &  0.002 & 0.88  \\
140 & 150 & 0.080 & 0.012 & 0.007 &  0.021 &  0.002 & 0.92  \\
150 & 160 & 0.117 & 0.013 & 0.006 &  0.021 &  0.002 & 0.95  \\
160 & 170 & 0.129 & 0.011 & 0.005 &  0.027 &  0.002 & 0.95  \\
170 & 180 & 0.108 & 0.012 & 0.007 &  0.026 &  0.002 & 0.94  \\
\hline
\multicolumn{8}{|c|}{\strut $10 < Q^2 < 30$ GeV$^2$ }  \\
\hline
0 & 90    & 0.004  &  0.001  &  0.001  & 0.000 &  0.001   & 0.68  \\
90 & 130  &  0.023 &   0.003 &  0.002 & 0.001 &  0.001  & 0.78  \\
130 & 140 &  0.042 &   0.010 &  0.007 & 0.003 &  0.001  & 0.79  \\
140 & 150 &  0.047 &   0.009 &  0.005 & 0.004 &  0.001  & 0.85  \\
150 & 160 &  0.057 &   0.010 &  0.003 & 0.005 &  0.001  & 0.91  \\
160 & 170 &  0.079 &   0.009 &  0.004 & 0.007 &  0.001  & 0.93  \\
170 & 180 &  0.064 &   0.009 &  0.005 & 0.007 &  0.001  & 0.93  \\
\hline
\multicolumn{8}{|c|}{\strut $30 \leq Q^2 < 350$ GeV$^2$ } \\
\hline
0 & 90    &  0.015  & 0.002  &  0.002  &  0.004 & 0.001  & 0.68  \\
90 & 130  &  0.040 &   0.004 &  0.003 & 0.011 &  0.001 & 0.83  \\
130 & 140 &  0.049 &   0.008 &  0.002 & 0.014 &  0.001 & 0.96  \\
140 & 150 &  0.030 &   0.008 &  0.001 & 0.017 &  0.002 & 0.99  \\
150 & 160 &  0.064 &   0.009 &  0.007 & 0.016 &  0.001 & 1.01  \\
160 & 170 &  0.046 &   0.007 &  0.005 & 0.020 &  0.002 & 1.01  \\
170 & 180 &  0.045 &   0.009 &  0.003 & 0.019 &  0.002 & 0.97  \\
\hline
\end{tabular}
%\end{adjustbox}
\caption{Measured differential cross-section $\frac{d\sigma}{d\Delta\phi}$. Details as in Table \ref{tab:dsdxgamma}. \label{tab:dsddphi}}
\end{table}

% Table 4===========
\begin{table}
\centering
%\begin{adjustbox}{width=1\textwidth}
\begin{tabular}{|l @{ -- } r | 
c @{${}\pm{}$} 
c @{(stat.)${}\pm{}$} 
c @{(sys.)\hspace{\tabcolsep}} |
c @{${}\pm{}$} 
c @{(stat.)\hspace{\tabcolsep}} |
c @{\hspace{\tabcolsep}} 
|}
\hline
\multicolumn{2}{|c|}{\begin{tabular}{c}\strutb$\Delta\eta$ \\range\\ \end{tabular}} 
& \multicolumn{3}{ c|}{ {\large $\frac{d\sigma}{d\Delta\eta}$   ($\mathrm{pb}$)}} 
& \multicolumn{2}{ c|}{ {\large $\frac{\dsigLL}{d\Delta\eta}$ ($\mathrm{pb}$)}} 
& \multicolumn{1}{c|}{\begin{tabular}{c}had. \\cor.\\ \end{tabular}} \\

\hline
\hline
\multicolumn{8}{|c|}{\strut $10 < Q^2 < 350$ GeV$^2$ } \\
\hline
--2.2 & --1.5 &  0.32 & 0.08 &   0.05 &  0.01 &   0.01 & 0.76 \\
--1.5 & --0.8 &  1.41 & 0.15 &   0.14 &  0.06 &   0.01 & 0.66  \\
--0.8 & --0.1 &  2.38 & 0.22 &   0.21 &  0.21 &   0.02 & 0.74  \\
--0.1 & 0.6   &  3.36 & 0.27 &   0.23 &  0.45 &   0.03 & 0.87  \\
0.6 & 1.3     &  3.88 & 0.28 &   0.22 &  0.87 &   0.04 & 1.04  \\
1.3 & 2.0     &  1.88 & 0.21 &   0.12 &  0.92 &   0.04 & 1.11  \\
\hline
\multicolumn{8}{|c|}{\strut $10 < Q^2 < 30$ GeV$^2$ }  \\
\hline
--2.2 & --1.5 &  0.14  &   0.05  &  0.03  & 0.00  &  0.01   & 0.63  \\
--1.5 & --0.8 &  0.51  &   0.12  &  0.04  & 0.00  &  0.01   & 0.68  \\
--0.8 & --0.1 &  1.16  &   0.15  &  0.09  & 0.04  &  0.01   & 0.77  \\
--0.1 & 0.6   &  1.70  &   0.19  &  0.15  & 0.08  &  0.01   & 0.90  \\
0.6 & 1.3     &  1.67  &   0.19  &  0.13  & 0.14  &  0.02   & 1.08  \\
1.3 & 2.0     &  0.71  &   0.13  &  0.07  & 0.13  &  0.02   & 1.07  \\

\hline
\multicolumn{8}{|c|}{\strut $30 \leq Q^2 < 350$ GeV$^2$ }  \\\hline

--2.2 & --1.5 &  0.20  &   0.07   &   0.03  & 0.00  &  0.01 & 0.83        \\
--1.5 & --0.8 &  0.86  &   0.09   &   0.09  & 0.05  &  0.01  & 0.65         \\
--0.8 & --0.1 &  1.25  &   0.16   &   0.13  & 0.16  &  0.02 & 0.72         \\
--0.1 & 0.6   & 1.68   &   0.19   &   0.08  & 0.37  &  0.03 & 0.85         \\
0.6 & 1.3     & 2.23   &   0.22   &   0.19  & 0.72  &  0.04 & 1.02         \\
1.3 & 2.0     & 1.16   &   0.16   &   0.06  & 0.80  &  0.04 & 1.14         \\

\hline

\end{tabular}
%\end{adjustbox}
\caption{Measured differential cross-section $\frac{d\sigma}{d\Delta\eta}$. Details as in Table \ref{tab:dsdxgamma}. \label{tab:dsddeta}}
\end{table}

% Table 5===========
\begin{table}
\centering
%\begin{adjustbox}{width=1\textwidth}
\begin{tabular}{|l @{ -- } r | 
c @{${}\pm{}$} 
c @{(stat.)${}\pm{}$} 
c @{(sys.)\hspace{\tabcolsep}} |
c @{${}\pm{}$} 
c @{(stat.)\hspace{\tabcolsep}} |
c @{\hspace{\tabcolsep}} 
|}
\hline
\multicolumn{2}{|c|}{\begin{tabular}{c}\strutb\Dphiegamma \\range \\ (deg)\\ \end{tabular}} 
& \multicolumn{3}{ c|}{ {\large $\frac{d\sigma}{\Dphiegamma}   (\mathrm{pb/deg}$)}} 
& \multicolumn{2}{ c|}{ {\large $\frac{\dsigLL}{d\Dphiegamma} (\mathrm{pb/deg}$)}} 
& \multicolumn{1}{c|}{\begin{tabular}{c}had.\\ cor.\\ \end{tabular}} \\

\hline\hline
\multicolumn{8}{|c|}{\strut $10 < Q^2 < 350$ GeV$^2$ }  \\
\hline
0 & 45    &  0.025  &    0.003  &   0.002  &     0.009  &   0.001 & 0.95  \\
45 & 80   &  0.042  &    0.004  &   0.003  &     0.010  &   0.001 & 0.94  \\
80 & 110  &  0.047  &    0.004  &   0.003  &     0.010  &   0.001 & 0.92  \\
110 & 135 &  0.068  &    0.006  &   0.006  &     0.012  &   0.001 & 0.85  \\
135 & 155 &  0.093  &    0.009  &   0.007  &     0.015  &   0.001 & 0.79  \\
155 & 180 &  0.085  &    0.008  &   0.008  &     0.013  &   0.001 & 0.73  \\
\hline
\multicolumn{8}{|c|}{\strut $10 < Q^2 < 30$ GeV$^2$ }  \\
\hline
0 & 45    &  0.013  & 0.002  &   0.002  &  0.002  &   0.001  &0.95   \\
45 & 80   &  0.018  & 0.003  &   0.001  &  0.002  &   0.001  &0.94   \\
80 & 110  &  0.024  & 0.003  &   0.002  &  0.001  &   0.001  & 0.91  \\
110 & 135 &  0.033  & 0.005  &   0.002  &  0.002  &   0.001  & 0.85  \\
135 & 155 &  0.031  & 0.006  &   0.002  &  0.001  &   0.001  &0.78   \\
155 & 180 &  0.038  & 0.005  &   0.004  &  0.002  &   0.001  &0.80   \\

\hline
\multicolumn{8}{|c|}{\strut $30 \leq Q^2 < 350$ GeV$^2$ }  \\
\hline

0 & 45    &  0.012  & 0.002  &   0.001  &  0.007  &   0.001  & 0.95  \\
45 & 80   &  0.024  & 0.002  &   0.002  &  0.009  &   0.001  & 0.95  \\
80 & 110  &  0.023  & 0.003  &   0.002  &  0.009  &   0.001  & 0.93  \\
110 & 135 &  0.036  & 0.004  &   0.003  &  0.010  &   0.001  &0.86   \\
135 & 155 &  0.063  & 0.007  &   0.005  &  0.014  &   0.001  &0.80   \\
155 & 180 &  0.047  & 0.006  &   0.004  &  0.011  &   0.001  &0.70   \\

\hline
\end{tabular}
%\end{adjustbox}
\caption{Measured differential cross-section
$\frac{d\sigma}{d\Dphiegamma}$. Details as in Table
\ref{tab:dsdxgamma}. 
}
\label{tab:dsddphi_e_ph}
\end{table}

% Table 6===========
\begin{table}
\centering
%\begin{adjustbox}{width=1\textwidth}
\begin{tabular}{|l @{ -- } r | 
c @{${}\pm{}$} 
c @{(stat.)${}\pm{}$} 
c @{(sys.)\hspace{\tabcolsep}} |
c @{${}\pm{}$} 
c @{(stat.)\hspace{\tabcolsep}} |
c @{\hspace{\tabcolsep}} 
|}
\hline
\multicolumn{2}{|c|}{\begin{tabular}{c}\strutb$\Detaegamma$ \\range\\ \end{tabular}} 
& \multicolumn{3}{ c|}{ {\large $\frac{d\sigma}{d\Detaegamma}   (\mathrm{pb}$)}} 
& \multicolumn{2}{ c|}{ {\large $\frac{\dsigLL}{d\Detaegamma} (\mathrm{pb}$)}} 
& \multicolumn{1}{c|}{\begin{tabular}{c}had. \\cor.\end{tabular}} \\

\hline\hline
\multicolumn{8}{|c|}{\strut $10 <  Q^2 < 350$ GeV$^2$ }  \\
\hline
--3.6 & --3.0  & 0.94 &  0.21 &   0.12 &  0.02 &   0.01 & 0.80  \\
--3.0 & --2.4  & 3.57 &  0.30 &   0.30 &  0.08 &   0.01 & 0.82  \\
--2.4 & --1.8  & 5.44 &  0.36 &   0.45 &  0.45 &   0.03 & 0.83  \\
--1.8 & --1.2  & 3.79 &  0.31 &   0.26 &  1.33 &   0.05 & 0.85  \\
--1.2 & --0.6  & 1.90 &  0.21 &   0.11 &  1.24 &   0.05 & 0.89  \\
\hline
\multicolumn{8}{|c|}{\strut $10 < Q^2 < 30$ GeV$^2$ }  \\
\hline
--3.6 & --3.0 &  0.93  &   0.21  &  0.12  & 0.02  &  0.01   & 0.81  \\
--3.0 & --2.4 &  2.60  &   0.25  &  0.19  & 0.06  &  0.01   &0.85   \\
--2.4 & --1.8 &  2.69  &   0.25  &  0.19  & 0.22  &  0.02   &0.89   \\
--1.8 & --1.2 &  0.86  &   0.15  &  0.07  & 0.19  &  0.02   &0.92   \\

\hline
\multicolumn{8}{|c|}{\strut $30 \leq Q^2 < 350$ GeV$^2$ }  \\
\hline
--3.0 & --2.4 &  1.00  &   0.17  &  0.11  & 0.02  &  0.01   &0.77   \\
--2.4 & --1.8 &  2.72  &   0.26  &  0.25  & 0.23  &  0.02   &0.80   \\
--1.8 & --1.2 &  3.00  &   0.27  &  0.18  & 1.14  &  0.05   &0.84   \\
--1.2 & -0.6  &  1.90  &   0.21  &  0.11  & 1.24  &  0.05   &0.89   \\
\hline
\end{tabular}
%\end{adjustbox}
\caption{Measured differential cross-section
$\frac{d\sigma}{d\Detaegamma}$. Details as in Table
\ref{tab:dsdxgamma}.
}
\label{tab:dsddeta_e_ph}
\end{table}

%------------------------------------------------------------------------------
%       Figures
%------------------------------------------------------------------------------

\newpage
%Figure 1

% figure 2
\newpage
%Figure 3

%Figure 4

%%%%%%%%%% Figure 5

%Figure 6

%-----------------Figure 7 - comparison to theory
\begin{figure}[p]
\vfill
\begin{center}
\Large
\textbf{ZEUS }
\normalsize
\mbox{
\hspace*{-0.03\textwidth}
\includegraphics[width=0.53\textwidth]{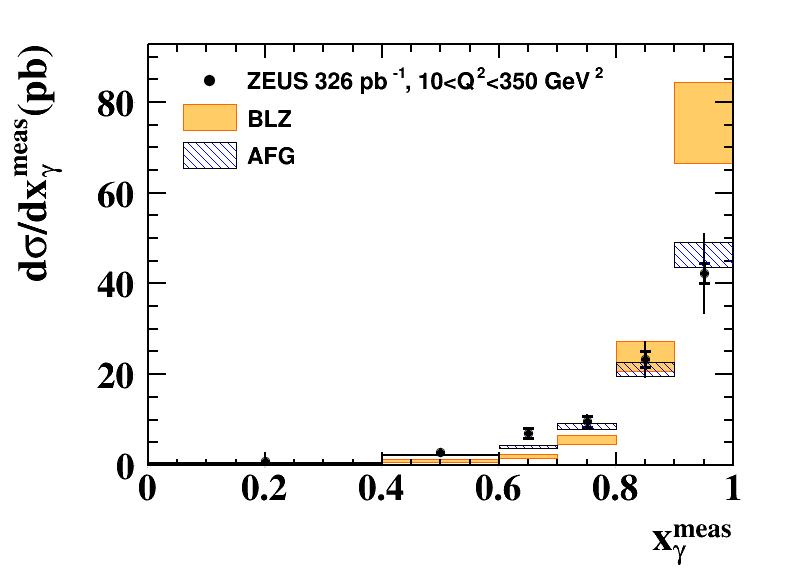}
\hspace*{-0.01\textwidth}
\includegraphics[width=0.53\textwidth]{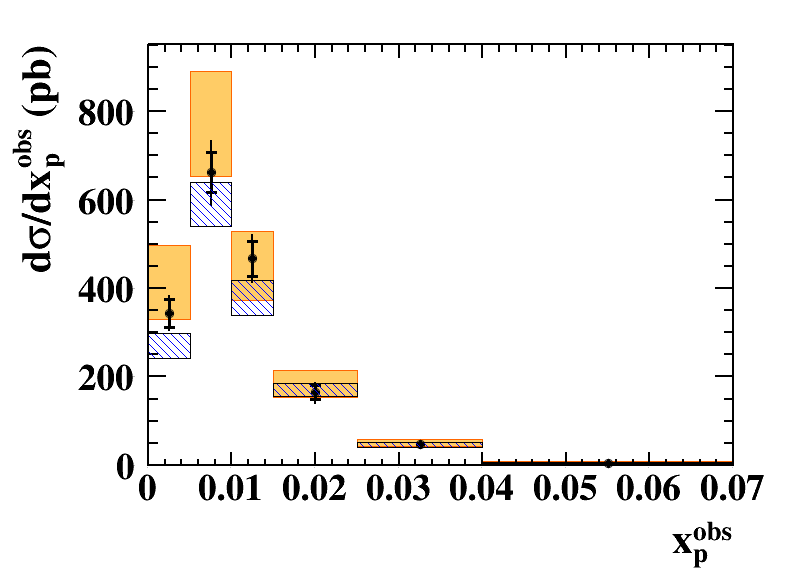}
}\\[-0.04\textwidth]
\hspace*{0.1\textwidth}(a)\hspace*{0.48\textwidth}(b)\\[0.04\textwidth]
\mbox{
\hspace*{-0.03\textwidth}
\includegraphics[width=0.53\textwidth]{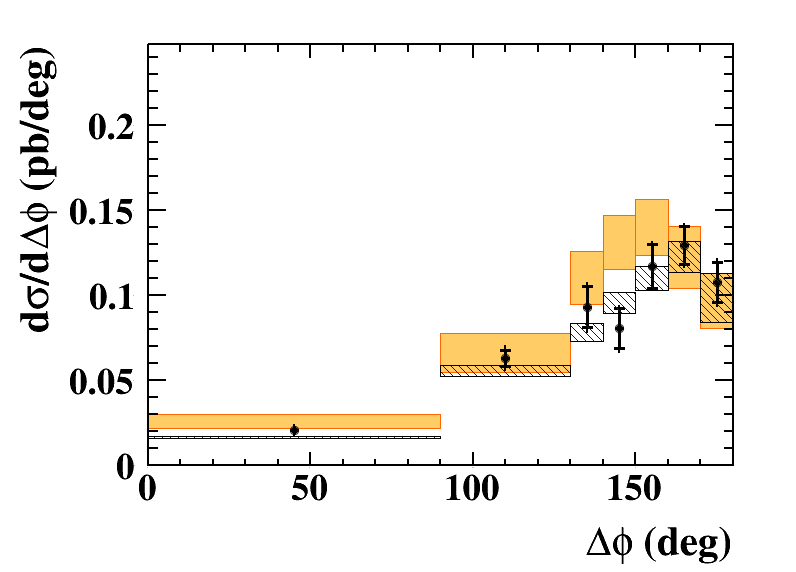}
\hspace*{-0.01\textwidth}
\includegraphics[width=0.53\textwidth]{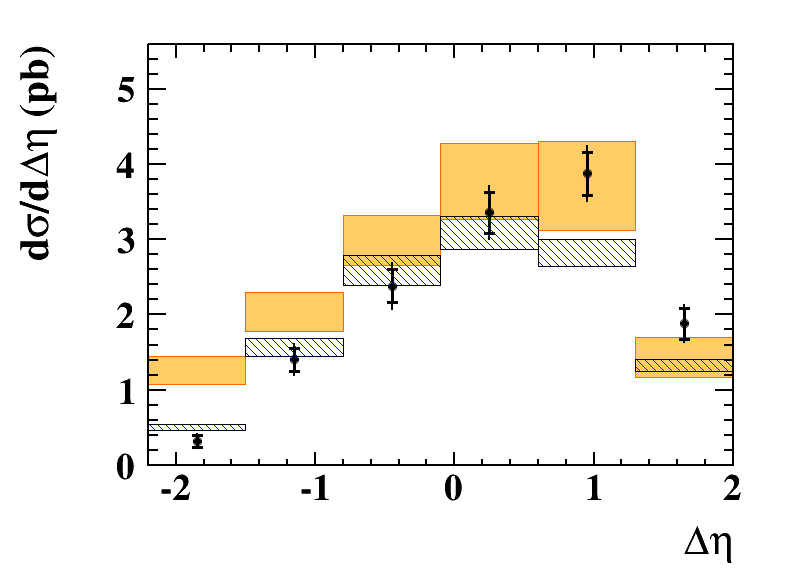}
}\\[-0.04\textwidth]
\hspace*{0.1\textwidth}(c)\hspace*{0.48\textwidth}(d)\\[0.04\textwidth]
\mbox{
\hspace*{-0.03\textwidth}
\includegraphics[width=0.53\textwidth]{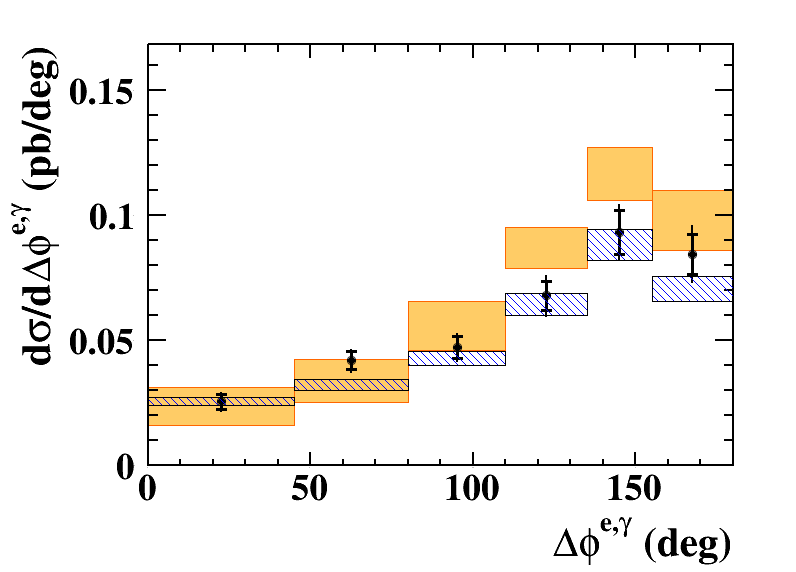}
\hspace*{-0.01\textwidth}
\includegraphics[width=0.53\textwidth]{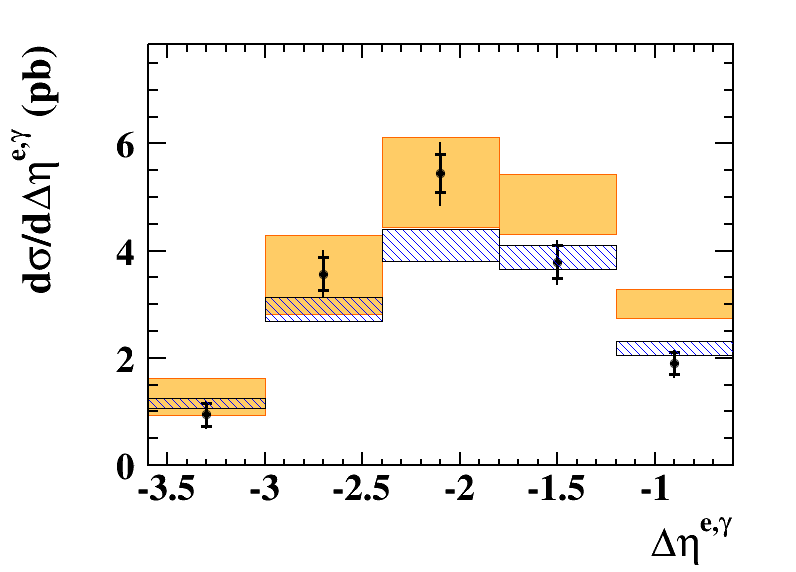}
}\\[-0.04\textwidth]
\hspace*{0.1\textwidth}(e)\hspace*{0.48\textwidth}(f)\\[0.04\textwidth]
\end{center}
\caption{\small Differential cross sections for selected variables in
the full $Q^2$ range $10<Q^2 < 350$ \gev$^2$:
as in Fig.~\ref{fig:xsec1}.  Theoretical
predictions from Aurenche \etal (AFG) and Baranov \etal
(BLZ) are shown, with scale uncertainties indicated by the 
bands.}
\label{fig:xsec6}
\end{figure}
\vfill

%--------------Figure 8 - comparison to theory 10-30
\begin{figure}[p]
\vfill
\begin{center}

\Large
\textbf{ZEUS }
\normalsize
\mbox{
\hspace*{-0.05\textwidth}
\includegraphics[width=0.54\textwidth]{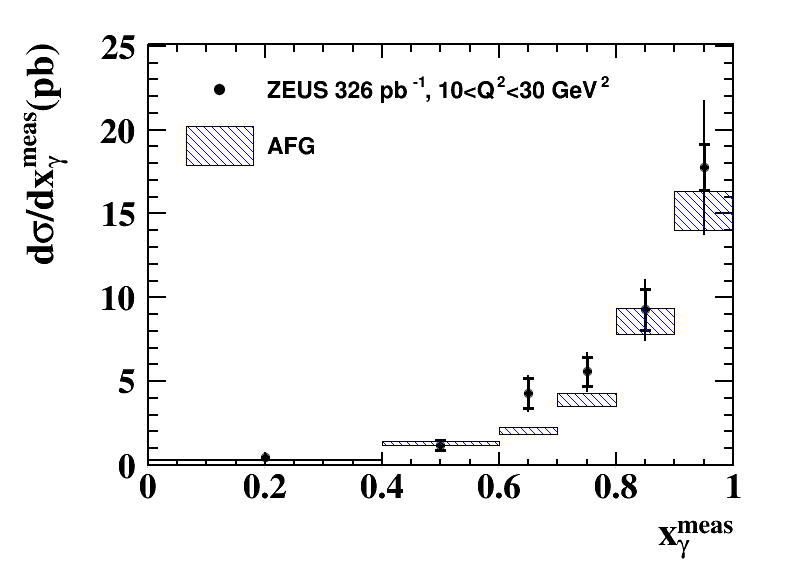}
\hspace*{-0.04\textwidth}
\includegraphics[width=0.54\textwidth]{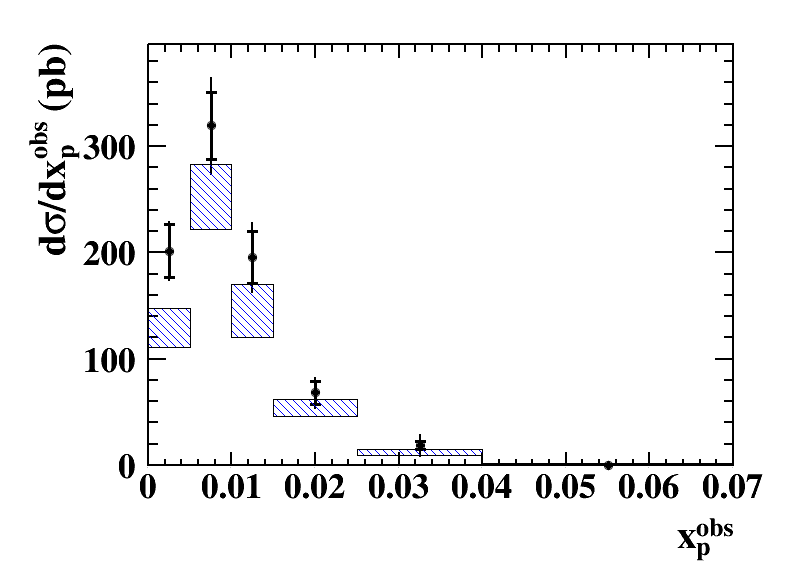}
}\\[-0.04\textwidth]
\hspace*{0.1\textwidth}(a)\hspace*{0.48\textwidth}(b)\\[0.04\textwidth]
\mbox{
\hspace*{-0.05\textwidth}
\includegraphics[width=0.54\textwidth]{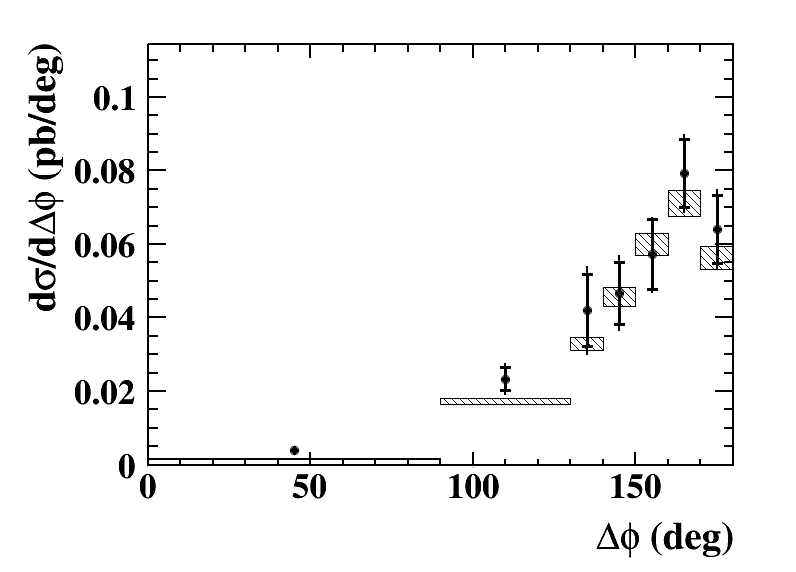}
\hspace*{-0.04\textwidth}
\includegraphics[width=0.54\textwidth]{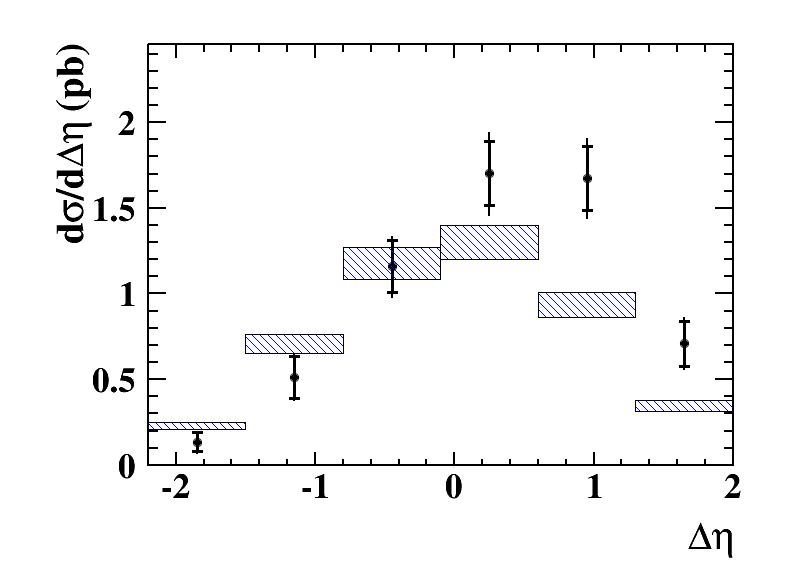}
}\\[-0.04\textwidth]
\hspace*{0.1\textwidth}(c)\hspace*{0.48\textwidth}(d)\\[0.04\textwidth]
\mbox{
\hspace*{-0.03\textwidth}
\includegraphics[width=0.53\textwidth]{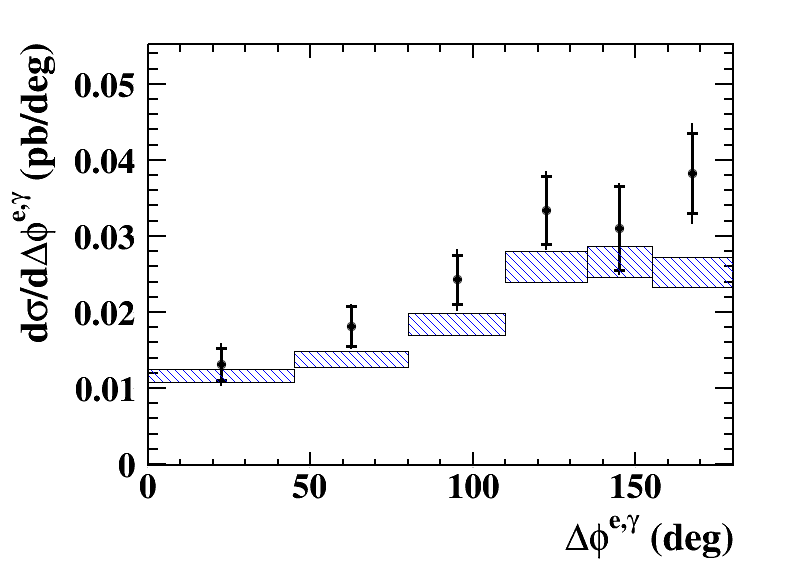}
\hspace*{-0.01\textwidth}
\includegraphics[width=0.53\textwidth]{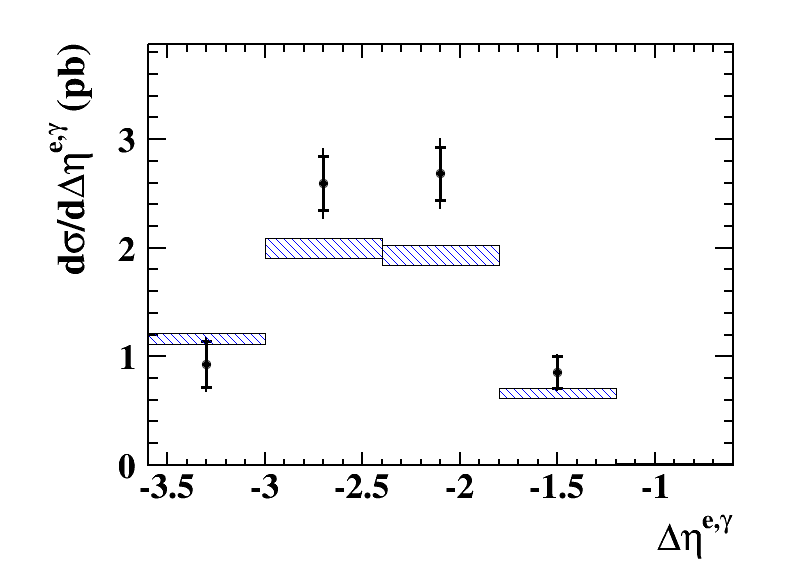}
}\\[-0.04\textwidth]
\hspace*{0.1\textwidth}(e)\hspace*{0.48\textwidth}(f)\\[0.04\textwidth]

\end{center}
\caption{\small Differential cross sections for selected variables in
the region $10<Q^2 < 30$ \gev$^2$  
as in Figs.~\ref{fig:xsec4},~\ref{fig:xsec5}.  Theoretical 
predictions from Aurenche \etal (AFG) are shown, with associated
uncertainties indicated by the bands.}
\label{fig:xsec7}
\vfill
\end{figure}

%Figure 9 - comparison to theory 30>350
\begin{figure}[p]
\vfill
\begin{center}

\Large
\textbf{ZEUS }
\normalsize
\mbox{
\hspace*{-0.05\textwidth}
\includegraphics[width=0.54\textwidth]{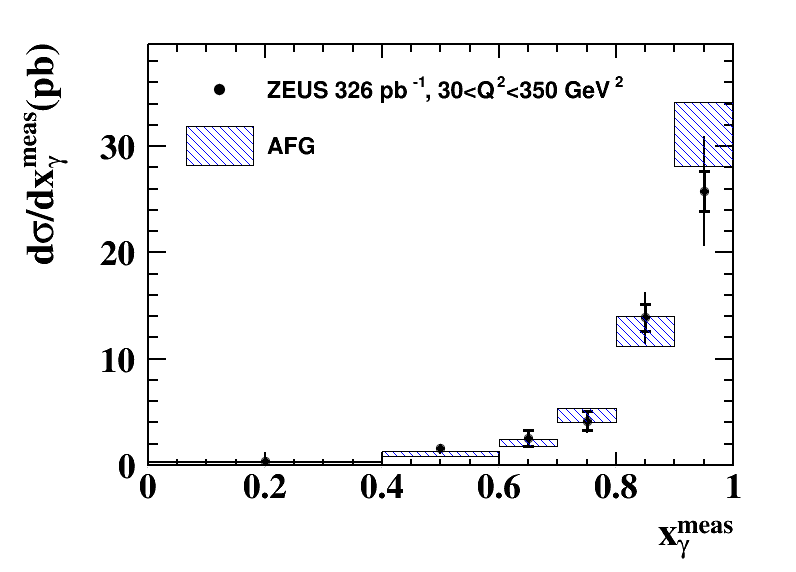}
\hspace*{-0.04\textwidth}
\includegraphics[width=0.54\textwidth]{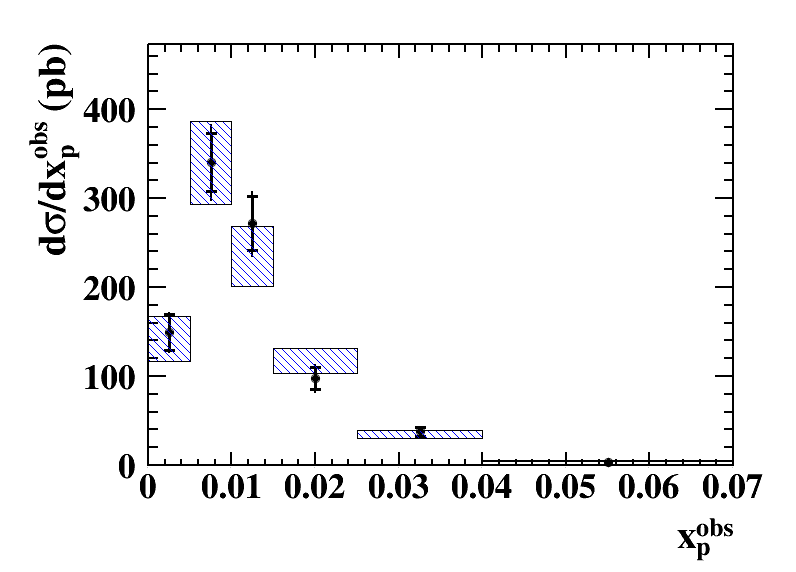}
}\\[-0.04\textwidth]
\hspace*{0.1\textwidth}(a)\hspace*{0.48\textwidth}(b)\\[0.04\textwidth]
\mbox{
\hspace*{-0.05\textwidth}
\includegraphics[width=0.54\textwidth]{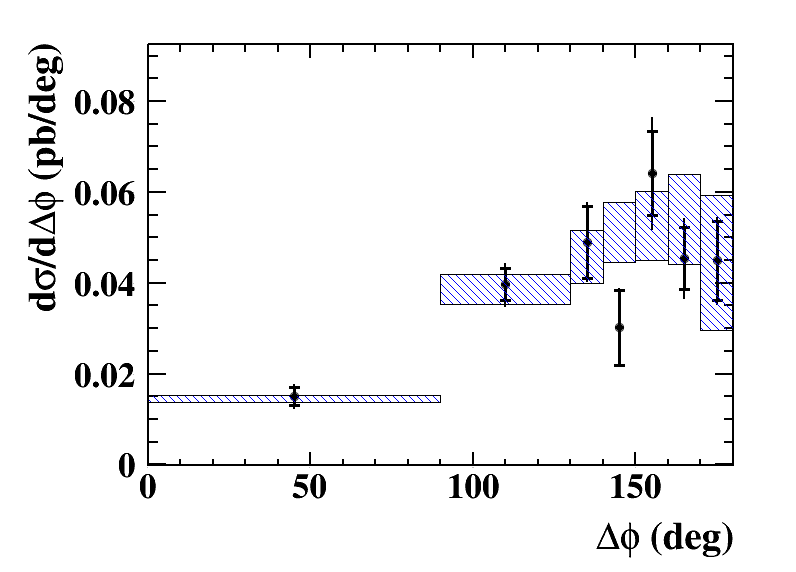}
\hspace*{-0.04\textwidth}
\includegraphics[width=0.54\textwidth]{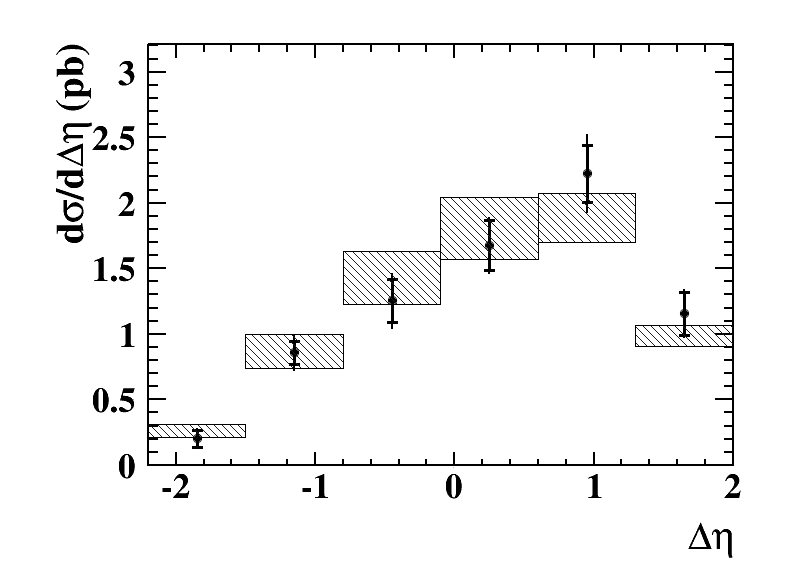}
}\\[-0.04\textwidth]
\hspace*{0.1\textwidth}(e)\hspace*{0.48\textwidth}(f)\\[0.04\textwidth]
\mbox{
\hspace*{-0.03\textwidth}
\includegraphics[width=0.53\textwidth]{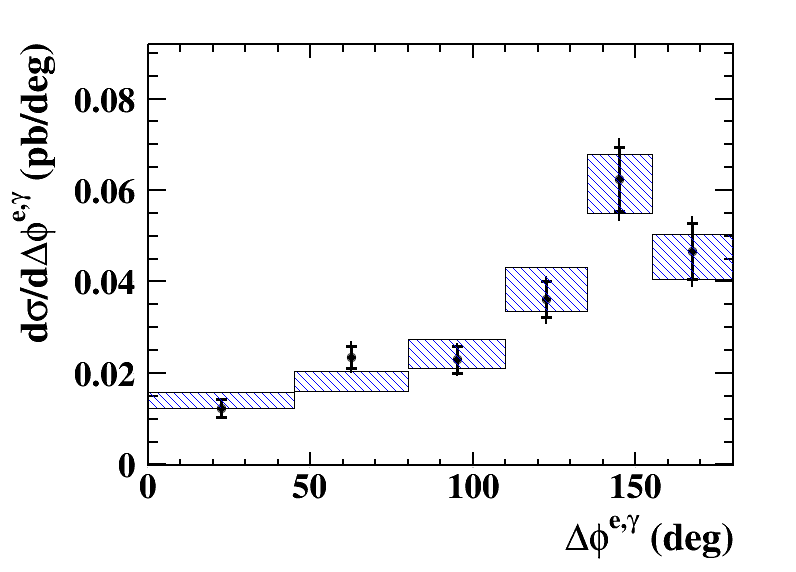}
\hspace*{-0.01\textwidth}
\includegraphics[width=0.53\textwidth]{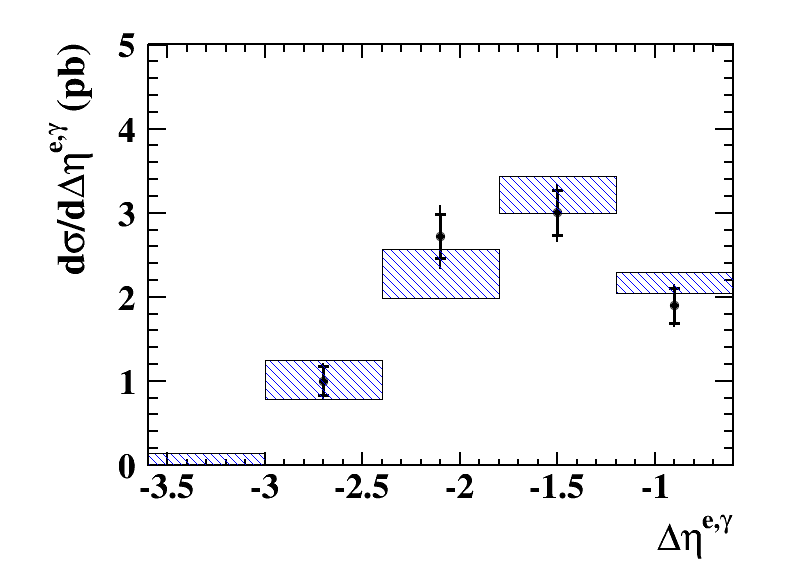}
}\\[-0.04\textwidth]
\hspace*{0.1\textwidth}(c)\hspace*{0.48\textwidth}(d)\\[0.04\textwidth]
\end{center}
\caption{\small Differential cross sections for selected variables in
the region $30 \le Q^2 < 350$ \gev$^2$ 
as in Figs.~\ref{fig:xsec4},~\ref{fig:xsec5}.  Theoretical 
predictions from Aurenche \etal (AFG) are shown, with associated
uncertainties indicated by the bands.}
\label{fig:xsec8}
\vfill
\end{figure}

\end{document}